# A Unified Theory of Language

Robert Worden

Active Inference Institute, Davis, CA, USA

rpworden@me.com

Draft 3.0; August 2025

Abstract:

A unified theory of language combines a Bayesian cognitive linguistic model of language processing, with the proposal that language evolved by sexual selection for the display of intelligence. The theory accounts for the major facts of language, including its great speed and expressivity, and much data on language diversity, pragmatics, syntax and semantics.

The computational element of the theory is based on Construction Grammars. These give an account of the syntax and semantics of the world's languages, using constructions (feature structures) and unification. Two novel elements are added to construction grammars: (a) an account of language pragmatics, and (b) an account of fast, precise language learning.

Multi-modal constructions (gestures, words, sequences of words, or pragmatic interactions) are represented in the mind as graph-like feature structures. On hearing first few examples of any construction, people use slow general inference to understand it. From a few examples, the construction is learned as a feature structure, and thereafter is rapidly applied by unification. All aspects of language (phonology, syntax, semantics, and pragmatics) are seamlessly computed by fast unification; there is no semantic/pragmatic boundary. This accounts for the major puzzles of pragmatics, and for detailed pragmatic phenomena. Unification of constructions (and facts in the common ground) guarantees that listeners recover the speaker's intended meaning, by unifying the same constructions as the speaker.

Unification is Bayesian maximum likelihood pattern matching. This gives evolutionary continuity between language processing in the human brain, and Bayesian cognition in animal brains. The human difference arises from sexual selection for language abilities. This is the basis of our mind-reading abilities, our cooperation, self-esteem and emotions – the foundations of human culture and society.

(268 words)

## 1. Introduction

In recent years there has been huge progress in the scientific study of language. Can this progress be brought together in a unified theory of language? Such a theory would be assessed in four dimensions:

1. **Descriptive**: Does it describe the known features of language?
2. **Computational**: Does it give an account of how language is computed in the brain?
3. **Evolutionary**: Does it give an account of how language evolved in the human species?
4. **Unity**: Can these be explained by a unified theory, with economy of hypothesis?

Language is usually studied under three headings – syntax (grammar and the structure of sentences); semantics (what utterances mean), and pragmatics (how we converse). Under each heading, a vast amount of empirical data has been gathered about the world's languages; the descriptive challenge is huge. The challenges include the prodigious power of language – that we learn such huge vocabularies, to express an unlimited range of meanings, in rapid cooperative conversations. Another challenge is to account for the conscious experience of language.

The computational challenge is to find out what computations in the human brain support our prodigious language capability. While we know which regions of the brain are involved in language, we do not know how language is processed neurally at the micro level. There are working computational models of language processing, including language learning, at Marr's [1982] level 2. Can these models be combined in a unified model of how all language works in the brain?

There is also the evolutionary challenge – how did the complexity of today's language, in society and in individual brains, evolve from earlier animal cognition? Why is language so prodigious, and why is it unique to *Homo Sapiens*? Can we relate the modern language capability to quantitative biological selection pressures? (not just to say qualitatively that 'there was a selection pressure', but to know that the selection pressure was strong enough to lead to language as we know it, and occurred only in humans.) Can we relate the computational language capabilities of the human brain to pre-existing capabilities in animal brains?

Together, these three dimensions present a huge challenge for a scientific account of language. Can that challenge be met by a unified theory of language?

Remarkably, I believe that the answer now is 'Yes'. We can draw together existing work on language, with no radically new ingredients, to make a unified theory of language which answers all the main questions under headings (1) – (3). Building on the work of many researchers, a new language synthesis is possible. This paper describes the main elements of that synthesis.

There is one important omission in this synthesis, under heading (2) of computation. [Marr 1982] defined three levels at which brains can be studied – a Computational Level 1, describing the computational requirement which brains need to meet; a Level 2 of algorithms and data structures, which is an abstract definition of how brains compute (a form which can be simulated on modern computer hardware) ; and finally, a Level 3 of neural implementation, how brains do the computation using neurons. In the theory of this paper, language computations are described at Marr's levels 1 and 2. There is not yet any good description of how neurons in the brain process language.

To this you may respond: what about neural net models of language? What about Large Language Models? While LLMs have some prodigious and surprising capabilities, neural nets are not a useful biological model of human language, for three reasons: (a) they inhabit a world of words and links between words, with no concept of what the words mean; (b) neural nets learn languages so slowly (with training measured in 'epochs' of thousands of learning examples) that they cannot be a model of human language learning; and (c) they have no conscious awareness of language meanings.

This is a long paper, and it is long for a reason. Language is a huge phenomenon, deeply entwined with our lives. A unified theory of language has not yet been attempted; this attempt would not fit in a shorter paper. To convey the unity of the theory, without excessive cross-referencing between papers, this length is necessary. To give a sense of the landscape, you may want to see the foundations of the theory in the next section 2, then skip straight to section 15, which is a summary comparison of the theory with the evidence. Then you can decide whether or not to dig deeper.

## 2. Key Elements of the Unified Theory

The key elements of the unified theory of language are:

1. Cognitive Linguistics (Construction Grammars) for syntax and semantics
2. Bayesian cognition
3. Unification in construction grammars is a form of Bayesian Maximum Likelihood Inference
4. Bayesian Learning Theory supports fast learning of language constructions.
5. Language evolved by Sexual Selection for intention reading, high intelligence and social status
6. Language started in pragmatic interactions and gesture.
7. Construction Grammars are extended to describe conversational pragmatics
8. Language has seamless integration between pragmatics, semantics, and syntax
9. Language change happens by fast evolution of constructions
10. Conscious language understanding works by building conscious spatial models of meanings

Most of these elements are not new; the theory unites existing ideas in language and cognitive science:

1. **Construction Grammars**: The theory starts from a non-Chomskyan, non-generative approach to syntax and semantics, as developed in cognitive linguistics and construction grammars (surveyed in [Hoffmann & Trousdale 2013]; see also [Fillmore 1982, 1985; Langacker 1987,1991, 2008; Goldberg 1995; Croft 2001; Kay 2002; Bybee 2010]). These use graph-like constructions (feature structures) and the core operation of unification, as in unification-based grammars [Kaplan & Bresnan 1981; Gazdar et al 1985; Shieber 1986], to describe language syntax and semantics. There are working online computational models of construction grammars [Steels 2013, Worden 2022a]
2. **Bayesian Cognition**: The Bayesian revolution in cognitive science [Rao, Olshausen & Lewicki 2002; Friston, Kilner & Harrison 2006; Chater & Oaksford 2008; Friston 2010] proposes that animal brains do not work with logical certainty; instead, animals continually take bets on states the world, to choose the actions with the greatest expected fitness. Animal brains have evolved to incorporate Bayesian prior probabilities of states of the world, and they use Bayes' Theorem to modify those prior probabilities in the light of current sense data. The Bayesian model of the brain has been very successful, agreeing with data in many areas such as perception, foraging, and learning. As Bayes' theorem is the foundation of animal cognition, we expect it to be important in human cognition, including language.
3. **Unification is Bayesian pattern-matching**. This brings together (1) and (2). The feature structures of construction grammars are partial probabilistic representations of states of the world. When a feature structure has an information content of I bits, it represents a set of states of the world with probability of the order $2^{-I}$. The unification of two feature structures A and B (written as C = A U B) represents a set of states which are described by both A and B. The unification C has the smallest possible information content compatible with both A and B; so the set of states represented by C the highest possible probability compatible with A and B; it has the Bayesian maximum likelihood. Unification in construction grammars is a form of Bayesian maximum likelihood inference – used to infer the most likely interpretation of an utterance. This close identification between cognitive linguistics and Bayesian cognition [Worden 2022a] is at the heart of the unified theory of language
4. **Bayesian Learning Theory**: Animals learn the regularities in their habitats very fast – typically from only a few learning examples - and there is a Bayesian computational model of how they do it [Anderson 1990; Worden 1995, 2024c]. This model is defined at Marr's [1982] level 2, and it does not yet have a neural implementation. By contrast, neural models of learning (neural nets) [LeCun, Bengio & Hinton 2015] learn far too slowly to account for animal learning, or to be a model of language learning [Elman 1990]. The construction grammar model of syntax and semantics can be extended to a working Bayesian computational model of language learning [Worden 2022a], which describes how children learn syntax and semantics, from only a few learning examples for each word.
5. **Language Evolved Partly by Sexual Selection**: theories of language evolution (surveyed in [Christiansen & Kirby 2005]) do not account for two prominent features of human language: its prodigious power and speed, and the fact that no other species has any comparable ability [Számadó & Szathmáry 2006]. Both facts are simply explained by the hypothesis that sexual selection (for the display of mind-reading ability, a form of superior intelligence) has played a part in the evolution of language [Miller 2001; Worden 2022b] Sexual selection leads to highly exaggerated traits, which are unique to a species [Lande 1981; Maynard Smith 1982] – as language is.
6. **Pragmatics were the Start of Language**: The sexual selection hypothesis of language can be combined with proposals from [Sperber 2000, Carston 2017, Levinson 2006, 2019, 2025] that pragmatic exchanges evolved as an 'interaction engine' before the evolution of spoken language. The earliest stages of language evolution were: (a) displaying knowledge of another person's intentions, in a common ground of shared understanding; (b) cooperative multi-step 'intention games' for the mutual display of knowledge, with simple turns; (c) intention games with compound turns (d) spoken language. The pragmatic skills for (a) - (d) are learned as constructions, and applied by unification. Alongside these pragmatic skills, the issues of the human self and self-esteem (social status) became central in our minds [Worden 2024b].

7. **Language change happens by fast evolution of feature structures**: The whole of a language, including its pragmatics, syntax and semantics, is embodied in a large set of constructions. These are passed from generation to generation by fast Bayesian learning. Each feature structure reproduces and persists by this learning process [Worden 2002]. In this reproduction, feature structures are subject to strong selection pressures, depending on their usefulness in the social group, brevity, ambiguity, and learnability. The set of feature structure 'species' in a language can change rapidly over a few human generations. This is the mechanism of historic language change, and it is responsible for the great power and diversity of the world's languages.

8. **Extending Construction Grammars to describe Pragmatics**: When two people converse, they use a Common Ground [Stalnaker 1978, 2002, Tomasello 2003, 2009] of information that they know is known by both of them; and they collaborate in turns to add information to the common ground, extending or repairing the conversation. Information in the common ground can be represented as feature structures. Language understanding, including pragmatic enrichment, can be modelled as unification, using feature structures from the common ground (also called 'context') as well as from the words spoken. In this way, most pragmatic inferences are done by unification, using pragmatic feature structures which are learnt by fast Bayesian learning. Pragmatic inferences include fast 'Theory of Mind' inferences of the other peoples' thoughts and intentions – all done by unification. These inferences are required by both speakers and listeners, in a shared model of inference that underwrites successful communication.

9. **Complex spoken Language is a Seamless Extension of Pragmatics**: The same computational model underlies both pragmatic exchanges and complex spoken language: fast Bayesian learning of feature structures, and fast unification of feature structures for performance. This is a seamless model, in that the same Bayesian unification operation underpins fast turn taking, language production and comprehension, syntax, meaning and disambiguation. There is no semantics/pragmatics boundary, and no handover from a semantics module to pragmatics.

10. **Conscious Language Understanding**: Any theory of language must address the issue of consciousness: how are we consciously aware of language meanings? While theories of consciousness are currently in a rudimentary state, it is known that much conscious awareness consists of a three-dimensional conscious model of local space. In this model of space, imagination is weakly superimposed as a transparent 'overlay' on present sense-based reality. Imagination includes imagined language meanings. In this theory, concrete language meanings are consciously experienced as imagined spatial models, and more abstract meanings are experienced largely through spatial metaphor. This accounts for the ubiquity of spatial metaphor in language [Lakoff & Johnson 1980].

The unified theory spans all aspects of language, using diverse concepts. There is much material to cover, and this is not a short paper. It has three main parts:

1. Sections 3-9 describe the foundations of the theory in points (1) – (7) above. These are mainly overviews of existing work, except for section 8, describing some early pragmatic stages of language evolution.
2. Sections 10 - 13 are the main new elements of this work – showing how the theory addresses data on pragmatics (points (8) and (9)), using the formalism of construction grammars and unification.
3. Section 14 addresses consciousness of language, point (10). Section 15 circles back to make an overview comparison of the theory with the facts of language. Section 16 describes some implications of this theory for human nature, and section 17 concludes.

The theory aims to account for the main known facts of pragmatics, syntax and semantics, in a working unified model of language computation in the brain, consistent with human evolution. To see if the theory achieves this aim, one route is to start at the summary of evidence in section 15, then come back to other sections as necessary.

## 3. Construction Grammars

The view of language developed in this paper is not derived from Chomskyan Generative Grammar [Chomsky 1965, 1980, 1995; Chomsky & Lasnik 1993] but starts from the competing framework of Cognitive Linguistics, also known as Construction Grammar [Fillmore 1982, 1985; Langacker 1987,1991, 2008; Goldberg 1995; Croft 2001; Kay 2002; Bybee 2010; Hoffmann & Trousdale 2013]. There are several reasons for this choice, besides the core reason, that cognitive linguistics is cognitive; it is an attempt to relate language to other human cognitive abilities. Some other reasons are:

- While generative grammar regards language as divided into a core language module and peripheral language performance [Chomsky 1980], construction grammars do not make this split. They are as much interested in issues like idiom, metaphor and less structured utterances, as in some 'core' grammar.
- Construction grammar views a language as a collection of constructions. A construction is a graph-like feature structure, similar to those which may be found for other aspects of cognition. A feature structure has a phonological pole and a semantic pole [Goldberg 1995]; the 'semantic' pole may contain pragmatic information as well as traditional semantics; construction grammars are extensible to pragmatics.
- Cognitive linguistics has made a wide study of the world's languages.
- Construction grammars have been widely tested by building working computational implementations, starting with the early implementations of phrase structure grammars [Kaplan & Bresnan 1982; Gazdar et al 1985; Shieber 1986].
- The core computational operation of construction grammars is unification. Unification can be seen as a Bayesian maximum likelihood operation – [Worden 2022a], relating construction grammars to Bayesian theories of animal and human cognition.

A construction is a feature structure, which can be computationally represented (at Marr's [1982] level 2) as a directed acyclic graph (DAG), with information stored as slot values on the nodes of the graph. A construction (in its semantic pole) represents a set of situations in the world, with the parts of situations nested inside other parts. Many of the semantic structures of construction-based linguistics are penetratingly analysed in [Jackendoff 1983, 1991], and in [Talmy 2000].

There are several flavours of construction grammars, including Berkeley Construction Grammar (BCG) [Fillmore 2013], Sign-based Construction Grammar (SBCG) [Sag, Boas & Kay 2012, Michaelis 2013], Fluid Construction Grammar (FCG) [Steels 2013], Embedded Construction Grammar (ECG) [Bergen & Chang 2013], Radical Construction Grammar (RCG) [Croft 2001, 2013], and Cognitive Construction Grammar [Boas 2013; Goldberg 1995]. These flavours are marked more by their similarity, with differences of emphasis, than by any deep differences of principle [Goldberg 2013]. Perhaps the variant most closely related to this work is Radical Construction Grammar [Croft 2001, 2013], in which constructions do not have slots describing grammatical roles; all slots describe either phonological or semantic information. Fluid Construction Grammar (FCG) is a computational toolkit which can be used to implement any variant of construction grammar. It can be used online [Steels 2013] to demonstrate important features of any construction grammar.

Unification is an operation of pattern-matching two feature structures[1], so that the result contains both structures, without repeating the parts where they match. Successive unifications build up larger feature structures, adding information. This paper has an appendix, describing the mathematical properties of unification.

A key attribute of unification, as described in the appendix, is that **it can be used in two different ways**. A construction can be selected for its semantic pole, and unified to add the information in its phonological pole (as is done in language production); or selected for its phonological pole, and unified to add information from its semantic pole (as is done in language understanding). A complex utterance can be generated or understood by unifying the same set of feature structures (constructions) – approximately one construction per word. The order of unification in understanding can be the reverse of the order in production, with a mathematical guarantee that understanding will recover the same meaning as was the input for production.

Language production takes a semantic tree structure of some intended meaning, and by successive unifications, adds to it a serialization of word sounds. Language understanding starts from the word sounds, and unifies the same constructions in a different order,to extend the series of word sounds by adding a meaning tree.

This two processes complement each other, as can be expressed as a **Communication Guarantee**:

I. If a speaker has a meaning M, and this meaning can be expanded into sounds by constructions with feature structures A, B, C, D, with the order of unification ABCD, then on hearing those sounds a listener can unify the same feature structures in the reverse order DCBA, and this is guaranteed to recover the same meaning M.
II. If a speaker does not possess all the feature structures to express the full meaning M, she may still express a subset M' of the meaning, and the listener (provided she has the same feature structures) can recover the meaning M'.

---

[1] In all the following, recall that a construction in Construction Grammar is a special type of feature structure, with a phonetic pole and a semantic pole. A feature structure is a directed acyclic graph, with slots on its nodes denoting fixed or variable information values, such a gender, number, semantic role or tense.

This guarantee is an important element of the unfied theory. It can be proved from the mathematical properties of unification. The mathematical basis of the guarantee is described in the appendix to this paper, It can be seen working in the Fluid Construction Grammar online computational model [Steels 2013].

The communication guarantee can also be seen working, for a 40-word subset of English, in an online demonstration of language production, understanding and learning [Worden 2022a]. This demonstration is restricted to declarative language syntax and semantics. A key result of this paper is to expand the scope of the communication guarantee, to give a model of pragmatics.

To produce a single utterance, the speaker may need to selectively retrieve and unify several constructions (typically about one construction for each word), from a lexicon of many thousands of constructions – retrieving constructions for many words within one or two seconds. Similarly, the listener needs to retrieve the same constructions from a large lexicon. This prodigious speed requires powerful associative retrieval mechanisms – for the speaker, selecting constructions by the semantics pole, and for the listener, selecting by the phonological pole.

In construction grammars, this retrieval is usually described by making loose reference to an **inheritance graph** of constructions [e.g. Goldberg 2013] - a wide, shallow graph whose node are feature structures (constructions), in which any construction may inherit from several ancestors. Selective retrieval returns constructions with strongest match first – starting with the most specific constructions which match.

Inheritance graph retrieval of constructions is usually not modelled in detail; and if it is modelled at all, the models are at Marr's (logical) level 2 – not his level 3 of neural implementation. Although it may be tempting to identify the inheritance graph with a network of neurons, nobody has worked out how to do this; and the mapping of constructions onto neurons is not likely to be simple.

All we know is that some such powerful associative retrieval is done somehow in the brain – retrieving and unifying constructions at a rate of several constructions per second. This all happens outside conscious awareness; only the final sounds or meaning come to awareness.

In summary, construction grammars give a working computational model of language syntax and semantics, including language learning. This has been tested in many languages. They do not yet address many pragmatic phenomena; that is addressed in later sections.

## 4. The Bayesian Theory of Cognition

It is now known that Bayesian maximum likelihood inference is a near-universal feature of animal cognition [Rao et al 2002; Chater & Oaksford 2008; Friston 2010].

It can be formally shown [e.g. Worden 1995, 2024c] that under uncertainty, Bayesian inference from sense data leads an animal to choose actions which give the greatest possible expected fitness; so that the evolution of animal brains is expected to converge towards Bayesian cognition, because it is maximum fitness cognition. This gives a strong theoretical reason to expect that animal cognition is near-Bayesian in character.

This expectation is confirmed empirically in many ways. A large amount of data, in fields as diverse as perception (of all kinds), food foraging, and animal learning, is consistent with the Bayesian brain hypothesis in all species [e.g. Anderson 1990; Krebs & Davies 1989].

So there are good evolutionary reasons to expect that, like the brains of all other species, the human brain relies on Bayesian cognition. Any theory of language which depends on Bayesian cognition will have the great advantage of evolutionary continuity between the human language capacity, and a core Bayesian capability of all animal brains. As we shall see in the next section, construction grammar is such a theory.

## 5. Construction Grammars as Bayesian Inference

In any language there is a large number of different feature structures, representing all the different words (and more complex constructions) in the language. Each construction has an information content, measured in bits and denoted by I. The semantic pole represents a set of possible situations in the world. For instance, the semantic pole of the construction for the word 'he' denotes a set of situations in which there is one male person present. Its phonetic pole represents any situation containing the word sound 'he'.

The set of situations represented by a feature structure has an approximate probability $2^{-I}$, where I is its information content - the total information content of all the slot values (in bits), got by summing over the slots in the construction. Two feature structures A and B are compatible (they can be unified) if the sets of situations that they represent overlap; they cannot be unified if the situations they represent are inconsistent. Their unification C = (A U B) represents the most likely set of situations which are described by both A and B; C contains both A and B as sub-structures, with the smallest possible total information content (maximum overlap of the slots of A and B); so the unification (A U B) represents the most likely set of situations compatible with both A and B.

Hence unification can be seen as a Bayesian maximum likelihood pattern matching operation. Given feature structures A and B, which each represent sets of situations, the result (A U B) represents the most likely set of situations which can be described by both A and B. Each unification is a Bayesian maximum likelihood inference, finding the most likely possible situation, subject to the constraints of the input feature structures.

In construction grammars, understanding of an utterance is done by unifying the feature structures for all the words (or other constructions) in it. This set of unifications both parses the utterance to reveal its phrase structure, and creates a meaning structure for the whole utterance. Since unification is a Bayesian maximum likelihood operation, the sequence of unifications finds the most likely possible parse of an utterance, and the most likely possible meaning. This is of great importance in making language robust, and able to handle ambiguities from many sources, such as background noise making words indistinct, or unknown words.

The Bayesian interpretation of unification in construction grammars is described in the appendix, also at [Worden 2022a], which has an online working demonstration of the robustness of Bayesian language processing.

## 6. Bayesian Learning

The Bayesian Theory of Cognition has successfully described many aspects of animal cognition, including learning [Anderson 1990; Worden 1995, 2024c]. This section describes its application to animal learning, and to human language learning.

The Bayesian analysis describes cognition at Marr's [1982] Level 1. That is, it describes the computation that animal brains are required to do (in terms of the inputs and outputs), without saying how that computation is done – either at a logical computational level (which is Marr's Level 2), or at a neural implementation level (which is Marr's level 3).

Bayesian cognition can be precisely described by a Requirement Equation [Worden 1995, 2024c] which states how any animal brain is required to choose which actions to take, given the animal's sense data, in order to give it the greatest possible fitness. The resulting equation incorporate Bayes' theorem applied to sense data. Because it gives maximum fitness, this is the form which animal brains evolve towards by natural selection. Brains act as if they make the Bayesian computation.

The Requirement Equation looks like Bayes' Theorem, with extra factors for the fitness payoffs of different actions. When the Requirement Equation is applied to learning, it gives a simple result: that an animal should learn about any regularity in its habitat, as soon as the evidence for that regularity, in its sense data, is statistically significant [Worden 1995, 2024c]. It should learn any rule about the world as soon as the rule is necessary to account for its experience. An animal can learn no slower than this, and no faster. In practice, the evidence for most regularities becomes statistically significant after only a few examples – so animals learn the regularities of their habitat very rapidly. Their survival depends on fast learning.

Animal learning has been tested over many years in experiments on associative conditioning. Anderson [1990] has compared the results of these experiments with Bayesian Learning Theory, and found good agreement. These experiments show that animals are Bayesian learners. Learning is so important for them that they have evolved to learn almost anything in their habitat as fast as it can be learned, and no faster. (if they tried to learn any faster, they would learn spurious regularities, from insufficient evidence).

In a typical associative conditioning experiment, an animal needs to learn a correlation between some set of features of its world (such as tones or lights) and something that matters to it, such as food or a reward. Animals learn these regularities very fast – typically from as few as 3-5 learning examples.

This fast learning agrees with Bayesian Learning theory, but does not agree with neural net models of learning [LeCun, Bengio & Hinton 2015]. Neural Net learning is typically measured in 'epochs' of many thousands of learning examples. No animal lives long enough to experience that many examples; most neural net learning would be useless to an animal. We do not yet know how fast Bayesian learning is neurally implemented in animal brains. We only know that it is done somehow, and it is nothing like current neural net models of learning.

Animal learning has other remarkable features:

1. Even small mammals and birds can learn large numbers of different rules – such as different rules of behaviour for many locations in their habitat.
2. Animals can learn complex rules, whose conditions depend on structured information, and which trigger complex behaviour – in both cases, more complex than just an unstructured set of attribute values – for instance, rules depending on sequences of events, or leading to sequences of actions. Complex learning is particularly seen in primates [e.g. Cheney & Seyfarth 1990; Worden 1996].
3. Animals can learn general rules and more specific rules [Anderson 1990], using the more specific rule only when it applies, and otherwise using the more general rule

These capabilities imply that some animals can learn rules with many of the complexities which are required for human language; learning and storing large numbers of rules, with the complexity of feature structures, with rapid associative retrieval of just the most appropriate rule out of hundreds or thousands of rules. Many of the computational learning capabilities needed for human language are found in other animal brains. We have no idea how these capabilities are neurally implemented, but they undoubtedly exist.

This brings us to human language learning. The well-known 'poverty of the stimulus' argument [Chomsky 1988] pre-dates the discoveries of Bayesian cognition and Bayesian learning; and it ignores the fact that a child, hearing some word for the first few times, has many clues (such as the meanings of other words in the utterance, speaker behaviour and other information in the common ground) to work out what it means, even if only approximately [Bloom 2000]. So it does not take long for a child to gather 3-5 learning examples to constrain the meaning and syntax of a new word – sufficient to learn the word by Bayesian learning.

So it is theoretically possible for children to learn the words and grammar of their native language at the high speeds we observe in children. This is more than just a possibility. [Worden 2022a] is a working online computational model of language production, comprehension and learning, in a radical construction grammar model of language (where all syntax is embodied in individual constructions). In the model, a child starts with no vocabulary and no grammatical knowledge. She hears random utterances made with a small vocabulary of about 40 words – and on some occasions (not all) can observe the utterance meaning and encode it as a feature structure. From this evidence, by Bayesian learning the child rapidly learns the meaning and syntactic constraints of the words – then being able both to produce and understand utterances with the words she has learnt. The online model also demonstrates a communication guarantee – that the listener can always reconstruct the speaker's intended meaning, provided they share the same set of language constructions.

In sum, if we assume a construction grammar model of language, and a Bayesian learning capability, then fast language acquisition is no longer a mystery. The learning model agrees with many empirical facts about child language learning, such as [Worden 1997, Baldwin & Tomasello 1998, Bloom 2000, Tomasello 2003, Bannard et al. 2009]

Having a powerful computational model of learning can make a big difference to our approach to many language phenomena, allowing our models to be 'learning-heavy', with thousands of learnt constructions. This applies particularly to pragmatics.

While the mathematical properties of feature structures and unification underpin the communication guarantee, another mathematical argument underpins a **learning guarantee**. It depends on the less-known operation of feature structure **generalization**. For two feature structures A and B, their generalization C = (A ∩ B) contains all the nodes and slot values that are in both A and B, and no others. The operation of generalization is complementary to unification, and it underpins language learning:

**Learning guarantee**:

> Suppose a child has learned the feature structures for words A, B, C…K, but has not yet learned word L. The child observes a small number of utterances, in which an adult uses some of the words A..K, and the word L. In these utterances, by observing the common ground the child is able to infer the meaning M that the adult is referring to, and encode it as a feature structure. By partially parsing the utterances using the words she has previously learnt, and using the inferred meaning M, the child constructs a small set of enriched learning examples – feature structures $E_1…E_3$. The generalization of these examples is the feature structure to be learnt: $L = (E_1 \cap E_2 \cap E_3)$. By this process, the feature structure L is faithfully propagated from adults to the child.

Together, the communication guarantee and the learning guarantee are the reason why language works, and is stable over the generations. By the communication guarantee, someone can produce an utterance by fast unification, or understand it by unifying the same feature structures in the reverse order to recover the speaker's meaning. The learning guarantee ensures that by learning, a group of speakers share the same set of feature structures which make this possible.

The mathematical basis of the learning guarantee is described in the appendix.

Both the communication guarantee and the learning guarantee can be seen working in the online demonstration at [Worden 2022a]. That model addresses phonology, syntax and declarative semantics, but it does not address pragmatics or context dependence. Section 9 - 13 of this paper describe how the same model can be extended to seamlessly accommodate pragmatics. First we need to address the evolution of the language capacity, and the proposal that pragmatics evolved before spoken language.

One obvious but important fact about language learning: while you can learn any word well enough to start using it from just a few learning examples, you never stop learning a word. Examples of any word that you encounter throughout your life continue to change its pragmatics, syntax and semantics. I have not explored this process.

## 7. Sexual Selection and the Evolution of Language

This section describes previous work on the evolution of language [Worden 2022b, 2024a, 2024b], which is extended and developed in this paper. The modification in this paper is the proposal, following [Origgi & Sperber 2000, Sperber 2000, Carston 2017, Levinson 2023, 2025], that pragmatic ability evolved as the first stage of language.

Almost all discussions of language evolution such as [Christiansen & Kirby 2005] have focused on the role of natural selection in human evolution – arguing that the fitness benefits of complex language (for instance in hunting [Számadó 2010]) have been sufficient to drive the evolution of our large brains and language capability.

These accounts suffer from three major difficulties [Számadó & Szathmáry 2006]:

1. They give no account of the uniqueness of human language; they do not explain why great apes, which have evolved in very similar habitats to us, under similar selection pressures, have not evolved any similar capability – to even a very limited extent.
2. The claimed fitness benefits of language would also arise from a communication capability which is much simpler and slower – for instance, with a specialised vocabulary up to 100 words, simple syntax, short sentences, and slow speech – taking as long as minutes to communicate an idea. They do not account for the prodigious power and speed of modern language, which is over-engineered for any habitat-related purpose.
3. The claimed fitness benefits of language – such as coordinating hunts, grooming, or planning foraging [Christiansen & Kirby 2005] – are not sufficient to offset the extra costs of our large brains – a 20% extra food requirement, and additional risks at birth. In natural habitats, complex language is a net handicap.

All three difficulties are solved, if language evolved not by natural selection alone, but by a combination of natural selection and sexual selection [Miller 2001; Worden 2022b].

Sexual selection is very widespread in biology, and it accounts for much of the visible profusion and diversity in nature. While sexual selection has been known since Darwin, the modern mathematical understanding of sexual selection dates to Lande [1981] and Maynard Smith [1982]. Sexual selection is not a slow, incremental adaptation towards some optimally fit phenotype for the habitat (like the evolution of animal brains, which has led to Bayesian cognition). It requires a co-evolution of genes which may be expressed differently in the two sexes – in one sex, to **show off** some visible trait; and in the other sex, to **size up** the extent of that trait in a potential mate.

Sexual selection involves positive feedback between 'show-off' genes and 'size-up' genes. For instance, peacocks show off a large tail; and peahens size up large tails. This positive feedback leads to very large selection pressures, in a runaway positive feedback process. The selection pressures are large, because every animal has one overriding aim in life – to pass on its genes to the next generation. To achieve this goal, it needs both to survive to adulthood, and find a mate. Survival depends on incremental adaptations to changes in the habitat – small selection pressures at any time, because the species is typically already quite well adapted to its habitat. But survival is no use without reproduction, which is an all-or-nothing matter; so sexual selection creates much larger selection pressures.

Every animal has cognitive facilities for choosing a mate – to size up potential mates, look for the fittest genes to combine with their own genes in offspring. These genes have random fluctuations. Suppose in a population of peacocks and peahens, a slight majority or peahens have a slight preference for larger tails. Males with larger tales have more offspring. Females with a stronger preference for large tails have more male chicks with larger tails, so have more descendants in the second generation. This is a runaway positive feedback process – reinforcing the show-off genes to grow very large tails, and the size-up genes to strongly prefer large tails.

Meanwhile, the larger tails are a handicap [Zahavi 1975] – increasing food requirements, making the birds less agile and more prone to predation. Some handicaps are worth it, if they lead to more offspring. Eventually the species settles at a tradeoff point, where the incremental benefits of a larger tails in reproductive fitness are balanced by the losses in habitat fitness. This can be expressed mathematically [Lande 1981; Maynard Smith 1982]. The tradeoff point is far from the point of best habitat fitness; a handicap has been locked into the peacock species.

In peacocks, show-off genes are expressed in the male, and size-up (mate choice) genes are expressed in the female. A more appropriate comparison for human language is the mating rituals or dances of certain birds – which require performance by both sexes, in that both sexes have both show-off genes and size-up genes. Competence in dancing can be assessed by either sex, and is needed by both sexes. An example is shown in figure 1.

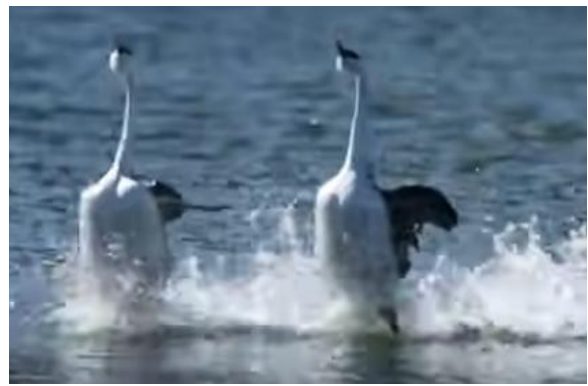

*Figure 1: The mating dance of western grebes*

In the mathematical models of Lande [1981] and Maynard Smith [1982], an important case is where mate choice depends on more than one variable – say, tail size and a tail colour. The species average is then a point moving in a plane of the two dimensions. Through sexual selection, the average moves away from the previous point of maximum fitness (the top of a fitness hill) but the direction in which it moves is arbitrary, and cannot be predicted or analysed; it depends only on some initial small random fluctuation, which is amplified by positive feedback.

This illustrates the general properties of sexual selection:

1. It depends on a positive feedback loop between show-off genes and size-up (mate choice) genes
2. The result is entirely species-specific, confined within one species.
3. The results are to some extent arbitrary; they cannot be predicted in detail
4. The results involve exaggeration of some traits, beyond the point where they serve any fitness purpose, to the point where they become a handicap.

These attributes of sexual selection relate closely to the three difficulties (1) – (3) in accounts of the evolution of language by natural selection. They suggest strongly that sexual selection was involved in the evolution of language.

Miller [2001] proposed that the greatly enlarged human brain is the result of sexual selection for increased intelligence, which is shown off and assessed in a wide variety of ways. Worden [2023] proposed that complex language is the fastest and most efficient way to show off intelligence, and that sexual selection was involved in the evolution of language. This solves the three major problems with accounts of language evolution by natural selection alone:

1. Language is unique to the human species, and nothing like it occurs in chimps, because chimps were not subject to the same sexual selection pressure. Sexual selection is always unique to a species.
2. Language is prodigiously faster and more powerful than it needs to be in a natural habitat, because its main purpose is competitive display. The competition has supercharged our language ability over thousands of generations.
3. In a natural habitat, language is a net handicap because it requires very large, metabolically expensive brains; but sexual selection often leads to handicaps, which can exist as long as they increase reproductive fitness.

The hypothesis that language evolved in part by sexual selection fits well with the facts of language, and with our evolutionary history. There is, as far as I know, no good reason to suppose that sexual selection was not involved in the evolution of language. I have not seen arguments against the hypothesis.

This paper refines that hypothesis in two ways:

1. It provides further detail on the time course of the evolution of language, proposing (following [Sperber 2000, Carston 2017, Levinson 2006, 2019, 2025]) that pragmatic exchanges by gesture evolved before spoken language.
2. It provides further detail about which facets of intelligence we need to display, in order to get a mate – particularly that we show off an ability to infer other peoples' intentions, and compete for status in our social groups.

I address here a possible objection which may have occurred to readers. They may feel uncomfortable that language, which is so central to our lives, could have arisen by sexual selection – which concerns only one aspect of our complex lives, which is important only during those few years when we are concerned with marriage and child-bearing. In response to this objection:

- Sexual selection gives rise to the strongest selection pressures, and those are the ones which shape the phenotype of any species, for all of an individual's lifetime.
- In many species, the competition to find a mate includes a competition to gain the highest possible status in the group; sexual competition is social competition. Competing for social status is central to our lives at any age, and language is a core part of the competition.
- The test of a theory of language is not the level of comfort you feel about its assumptions – it is how well the theory accounts for the facts of language. If the theory solves the three major problems above, that is more important than a comfort factor; it is what a theory is required to do.

This theory claims to offer a coherent account of the facts of language, both its evolutionary origins and its current state, including the main facts of pragmatics. Much work remains to be done; but please consider this theory a possible starting point.

## 8. Language Started in Pragmatic Interactions

This section extends the theory of language evolution, to describe the early stages in more detail. It builds on proposals by [Sperber 2000, Carston 2017, Levinson 2006, 2019, 2025] – that pragmatics came before spoken language, in both evolution and individual development.

Levinson calls this the 'Interaction Engine' hypothesis: that a package of pragmatic interaction skills (including joint attention, nested turn taking, and repair) evolved before spoken language, and serves as the scaffolding on which language evolved, and within which it can be learned.

He noted that: '*The objection may be made that this interactional ability is not one thing, but rather an assemblage of various talents and proclivities, with different phylogenetic origins and different ontogenetic patterns of development….*'. While not defending the interaction engine as a Fodorian mental module [Fodor 1983], Levinson called it a 'package' of abilities, and sketched a possible timeline for its evolution – with spoken language as the last stage in the timeline.

We can combine the sexual selection hypothesis of language evolution with the interaction engine hypothesis, to describe a possible timeline, which gives a reason for the unity of the language package. Its unity follows from the overriding need to win in a sexual selection game; that is the purpose of the interaction engine. The package of language talents is unified by that aim.

In this picture, the propensity which we initially needed to show off and size up, when selecting a sexual partner, is the ability to infer another person's intentions, in an 'intention game'. Later this expanded, from being a competition to exhibit just one attribute tied to reproductive success (intention reading), to be a broader competition for social status in the group – an 'intention and status game'.

The stages in this evolution could have occurred in several different orders, with overlaps. I shall describe one possible ordering, to bring out how various attributes of language and social behaviour could have emerged. I shall then discuss how much or little we can know about the actual ordering of these stages.

The illustrative ordering of stages is:

1. Single-turn displays of intention reading
2. Multi-turn intention games, with simple turns
3. Expressive gestures (mime)
4. Multi-turn games with compound turns
5. Syntax within turns; agency and social status
6. Use of the voice channel, and increased scope of meanings

Through this sequence, we shall see how the important attributes of language start and grow – attributes such as the common ground, mind-reading, turn structure, folk theories of the world, syntax, agency, shadow audiences, and assessing social status.

### Stage 1: Single-Turn Displays

The first stage may have happened up to three million years ago, around or before the first archaeological evidence of stone tools. To illustrate this stage, suppose I observe you returning from the woods carrying some sticks. For some reason, the cultural practice in our hominin group is to strip bark off the sticks. So I infer that you intend to strip bark; to do that, I know that you need a sharp stone; and I know where to find one. I pick up the stone, and offer it to you. That is all.

In this simple interaction, I **show off** to you that I can infer your intentions. You (and possibly others in the group) see this and **size up** my ability to infer your intentions. This is the classic show off/size up pairing of sexual selection. That increases my attractiveness as a potential mate, and increases my social status in the group.

In this first stage of sexual/social selection (as in the later stages) there is some practical benefit from our communication; it makes us both a tiny bit fitter, by reducing the time taken for some necessary daily activity. But the practical benefit is miniscule, compared to the benefit to both of us in reproductive fitness. Sexual selection is stronger than natural selection.

We compete with others in the group, in our ability to understand our common ground (the set of facts that you and I both know - about the current moment, our recent history, or group practices) [Stalnaker 2002; Tomasello 2003, 2009]; in our ability to infer one anothers' intentions from those facts; and our ability to show off that knowledge. At this stage, the runaway positive feedback of sexual selection for intention reading has just begun.

The first stage is shown in figure 2. In these diagrams, time goes from left to right.

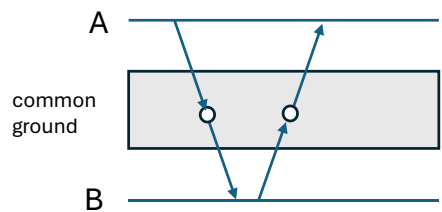

*Figure 2: First stage intention reading; B shows off, and then A sizes up*

Here, A initially does nothing new; then B shows off an intention-reading ability, which A then sizes up.

The knowledge used by B can be encoded as construction-like feature structures:

> X carries sticks => X intends to strip sticks
>
> X intends to strip sticks => X needs a sharp stone

X refers to any individual, including A. B infers that A needs a sharp stone, by unifying these two feature structures, using the fact 'A carries sticks' which is in the common ground between A and B. Even at this stage, this is a Theory of Mind inference – that a person A has a certain need, or intention – encoded as a feature structure. Mind-reading in conversation [Sperber & Wilson 2002] started very early.

In an encounter of the first stage, A planted a new fact in the common ground, but A did so unintentionally; B took the opportunity to plant another fact in the common ground, which scored a point in the group competition for mates. B showed off, and A sized up.

### Stage 2: Multi-turn game with simple turns

When the interaction becomes a multi-turn game, it gives both players multiple opportunities both to show off and to size up, as shown in figure 3.

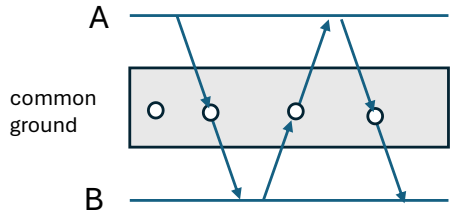

*Figure 3: Multi-turn intention game, with simple steps*

Consider again the first example, with one small difference. When carrying sticks, A does so **ostensively** – say, swinging the sticks from side to side - signalling to B (or to anyone else who is watching): I want you to know that I have got some sticks, and I intend to strip the bark off them.

B can respond positively, or so can anyone else who is quicker off the mark. The competition for speed of intention reading has begun (leading eventually to the 200 millisecond turn-taking of modern conversation [Levinson & Torreira 2015])

In each turn of the game, one player (e.g. A) deliberately plants one piece of information in the common ground; that is why they are 'simple' turns, with one fact added to the common ground per turn. This is the 'show off' element of the turn, and it is an ostensive communication: 'I have done this because I want you to work out my intention in doing it'. It is a test for B: 'can you work out why I did it?', leading to a subsequent sizing up of B by A: 'did B realise my intention?'. Initially, for B to work out what A is trying to communicate may require slow, conscious mental simulation and inference, from facts in the common ground; but as soon as B has observed the same move of the game a few times, B's knowledge of the move can be encoded as feature structures, applied by rapid, pre-conscious unification. B learns to react fast. A and B may develop codes between them: 'when I make this gesture, I mean….'; and these codes spread by learning across the group.

This multi-step intention game gives both players many more opportunities than the first single-step interactions, both to show off and to size up their intention reading. Players can make the game more complex, to better size up other players. The tools used for communication can include gestures, gaze, shared attention, facial expression, and simple vocalisations. Making them iconic and symbolic, rather than self-explanatory, is one way to make the game a more stringent test of learning and intelligence.

Even at this stage, some well-known features of pragmatic interactions emerge, as hominin groups develop their own group-specific games:

1. **Cooperation**: both players have an interest in extending the game, as it gives them more opportunities both to show off and to size up. Gricean cooperation [Grice 1989] emerges not as a rational way to achieve some practical objectives (achieving practical objectives is just a small side-effect), but for the more vital purpose of getting a mate.
2. **Turn-Taking**: A structure of clearly delineated turns [Levinson 1983] allows each player to show off and size up in a structured, learnable way; more so than in a free-for-all of interruptions, which would be harder to learn from, and has a stronger risk of breakdown. A player with over-active interrupting genes will not find partners in the game, or partners for reproduction.
3. **Complex and nested intentions**: As brains evolve to make people better players, they need to elaborate the game, to make it a more stringent test, to better size up other players. This may involve communicating more complex intentions, depending on a wider range of facts in the common ground; or conveying nested intentions within intentions [Levinson 1983, 2023]. It becomes a test of general intelligence.
4. **Repair**: Each player needs to test the other players as much as possible, by setting them progressively harder tests. This creates a risk that even a competent player will not 'get it', and the game will break down, which would result in a loss for both players. So there is a need for 'please repair' signals: 'what do you mean?'. [Levinson 1983] If player A repairs – conveys some information more explicitly – and B understands, the game can continue. More than that, B gets a learning example, to know in future what A means by some gesture, in the presence of other relevant facts in the common ground. Through repair, B can learn the feature structures for tricky cases.
5. **Greeting**: People will issue greetings – as an invitation to play the game and win sexual/social selection status points.
6. **Fast Turn-Taking:** If A issues an invitation to play the game – an ostensive communication - and several people are present, the first one to respond will get to play the game. In any turn of the game, players size up other players by the speed of their responses; so fast turn-taking [Levinson & Torreira 2015] is selected for, regardless of any practical benefit it may have.
7. **Slow conscious learning, fast pre-conscious performance**: Since the interaction game values

complexity, and it also values speed, players must find a way to understand complex intentions at speed. The solution is to learn feature structures – by a process of generalising across a few initial learning examples, which may require slow conscious inference or repair; followed by fast, pre-conscious unification of those feature structures.

8. **Relevance-based inference**: On first encountering an act of ostensive communication by player A, player B needs to infer what it means, based on a range of potentially relevant facts in the common ground [Sperber & Wilson 1986, 1995, 2012]. This may require slow conscious inference, sifting through facts in the common ground to assess their relevance; or if that fails, it may require repair by A. But then it provides a learning example for B, to learn the inference as a feature structure; after a few examples, B can know what A intends by fast unification – avoiding slow general inference.

9. **Turn Variety, and Turn Types**: Within each social group there will be a wide range of specific turns to be learned, making the game a more and more stringent test of intention reading (foreshadowing the great diversity of languages); but there will be a tendency in each group (and to a limited extent, across groups) for turns to be grouped into a few main turn types such as request, declaration, repair, and so on. These turn types will be encoded as feature structures high in the inheritance hierarchy. They are the forerunners of speech act types [Levinson 1983, 2023].

10. **Metaphor and Irony**: Various tropes common in speech may have their origins in this earlier stage of the intention game. For instance, there is a simple spatial metaphor between social status and physical height. A certain hand gesture may be made at different heights to convey the social status of someone being referred to. In a similar way, other graded scales, such as time, hunger, or physical strength, may be related metaphorically to physical space. Or, in the beginnings of irony, someone may convey an ironic attitude to a fact by making a hand gesture far away from his body – indicating 'that is what someone else thinks, not what I think'.

11. **Mental rehearsal of turns**: In order to make our turns in the intention game appear faster and more impressive, it was possible to consciously imagine some future turn, and use fast unification of feature structures to infer the consequences of such a move, in the eyes of some real or imagined 'shadow audience' – asking: 'If I make this gesture, what will person P think of it?' – where person P is the imagined shadow audience for the gesture.

Points (6) and (7) have particular importance as precursors to language. Already at this stage, both speed and power of intention reading are at a premium – they are selected for in players of the game. Conscious mental simulation and general-purpose inference would be too slow. The design solution is to use fast unification to reproduce the results of previous inferences; and fast learning to share a set of feature structures across a community. The communication guarantee, and the learning guarantee, are both needed to ensure that this works. (a pragmatic device, dependent on adding only one fact to the common ground, may still depend on several existing facts in the common ground)

Possibly before spoken language, the intention game places a premium both on fast learning, and fast performance – the ability to learn and retrieve just the right feature structure, selected from many stored feature structures, depending on diverse facts in the common ground; and the ability to share the same feature structures in a community. These are the abilities required for fluent spoken language.

### Stage 3: Expressive Hand Gestures

Stages 1 and 2 of the intention game can be carried out with whole-body gestures, perhaps supplemented by gaze, facial expression, and simple sounds. However, the human hand has evolved to be more controllable and potentially expressive than other primate hands. The driver for this evolution may have been the need to convey complex meanings – to mime complex actions, locations and things – in early intention games.

This evolution increased the number of available meanings; and because hand gestures can expressively mime, it gave listeners clues as to the intended meaning, and helped them consciously imagine the meaning. The need to keep up with this expansion of possible meanings drove both the evolution of expressive hands, and of clever brains to understand them.

Throughout this evolution, the common ground of shared meanings grew in complexity. The inference that 'fact X is in the common ground' is equivalent to the Theory of Mind inference that 'my partner in conversation knows fact X'. People learned an increasing range of feature structures encoding Theory of Mind knowledge for fast inference. Humanity developed a Fast Theory of Mind.

As the common ground grew more complex, it came to include a 'folk theory of the world' – a kind of encyclopaedic collection of all knowledge useful to the social group. This theory can be envisaged as a connected network of feature structures, making links between different concepts. The way into the folk theory is to learn the feature structures for the nascent words (or constructions) which are turns in the intention game. Proficiency in the folk theory becomes a determinant of success in the intention game. These folk theories of the world become group-specific, and undergo cultural evolution over short timescales. They are folk theories (assessed by their usefulness in the game) rather than true theories (assessed by their fit to reality).

Crucially, if the intention game had a mainly practical purpose – of somehow helping people to survive in the habitat – then we would expect the game to remain as simple as it can be, while fulfilling that purpose. However, the game becomes more and more complex, because its main purpose is to serve as a competitive test of peoples' intellects, in sexual selection. Through successive generations, people get smarter, so the test has to get harder.

**Stage 4: Multi-turn game with compound turns**

In this stage, there is still a clear structure of turn taking; but now, each gesture can convey an increasing range of meanings, make it useful to make gestures in sequence to convey complex meanings. Each player may in one turn place several facts or intentions in the common ground, implying that to infer his intentions, the other player needs to consider all those facts – possibly together with other facts already in the common ground, including facts and rules in the group's folk theory of the world. This is shown in figure 4.

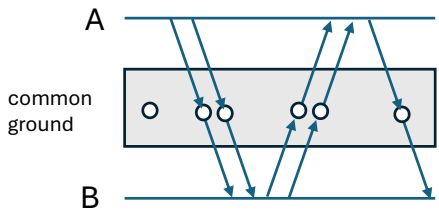

*Figure 4: Multi-turn intention game with compound turns*

This elaboration allows the game to become a yet more stringent test of intention reading. In any turn, a player needs to unify all the newly added facts in the common ground – put there by the other player in the latest turn – with any number of possibly relevant existing facts in the common ground.

The transition from stage 3 to stage 4 is the usual progression of sexual selection - as the attribute being selected for (in this case, intention reading) becomes more widespread and exaggerated in the population, so the job of sizing up that attribute requires ever more stringent tests. The game is further elaborated to make it harder.

**Stage 5: Simple Syntax: Agency and Status**

At this stage, the interaction game requires essentially all the capabilities needed for spoken language, except speech. Each single arrow in figure 4 is a construction – a feature structure – relating some existing facts in the common ground, and adding a new fact made by some player (e.g. a gesture) as a new fact in the common ground. For instance, A and B may have mutually understood gesture symbols for the meanings 'throw' and 'rock'. By making these two gestures in sequence, A communicates ostensively that he intends to throw a rock. To understand this intention, B needs to unify the two feature structures. The communication guarantee and the learning guarantee ensure that this process works – B understands A correctly.

Both the communication guarantee and the learning guarantee are needed – so that people can play the intention game by unifying the same shared feature structures, without resorting to slow conscious inference or slow repair. The required learning and inference use more complex feature structures, but still obey the mathematical principles which underpin the guarantees (as described in the appendix).

Figure 5 is an illustration of how simple syntax works.

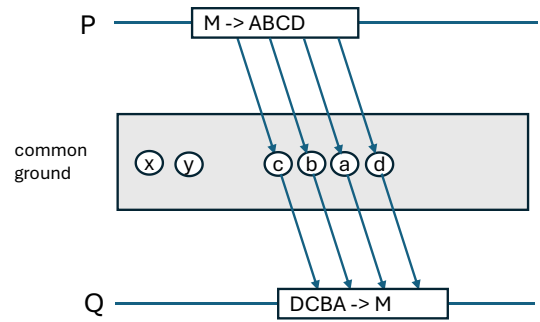

*Figure 5: One turn in the intention game*

This shows one turn in the intention game, in which P wishes to convey a meaning M. He has available a set of feature structures A, B, C.. whose 'phonetic' (gestural) poles are denoted by a, b, c…. By unifying these in the order ABCD (possibly unifying with pre-existing facts x and y in the common ground) P produces a series of gestures 'cbad' (in an order different from 'abcd', because of time ordering constraints in the feature structures). This creates a set of gesture facts c, b, a, d in the common ground. Q observes those facts, and has learned the same set of feature structures A, B, C…. By unifying these in the reverse order DCBA, Q can re-create P's original meaning M, because of the communication guarantee. Because of the learning guarantee, all speakers in a group learn the same set of feature structures.

At the end of this exchange, the fact M is in the common ground, because both A and B know it. In this way, simple syntax starts to emerge in the relative timings of hand gestures and other channels such as expression and gaze.

I note one prominent feature of language syntax, which may have emerged at this stage. The syntax of nearly all the world's languages has a common structure of Agent-Act or Agent-Act-Patient. This structure attributes agency to one of the participants; one participant is regarded as the 'agent', and is regarded as being responsible for the event taking place. Language need not have been like this; syntax might have developed to take a more neutral view of events – assigning people and things a large set of roles in an event, without giving special 'agent' responsibility to any one role.

The purpose of agency in language is to make people responsible for acts – so that status (in the form of blame or praise) can be assigned to people.

Biologically, agency does not exist in the animal world. A lion does not choose to hunt; it is simply impelled to hunt in certain circumstances. Evolution has created lions as a part of their ecology that inevitably hunt at certain times. Agency is a human innovation. Is it an innovation of the human mind (in that our minds have a new capacity for agency, for a kind of free will, denied to all other animals), or is it merely an innovation of human language?

If agency is merely an innovation of language, it serves a purpose there, playing a key role in sexual selection. By implying that a person is the agent in an event, language makes that person responsible for the event. Status values for the outcome can be assigned to the person, according to the norms of the social group, as embodied in its folk theory. Agency in language makes it possible to assign status points to people, changing their status as perceived by the group.

Agency expands the game, from being a straight intention-reading game, to an intention and status game. This change was a rubicon for mankind. It established the role of language in our minds, as not merely describing the world, but also as reasoning about our social status. This status-computing role of language now occupies our minds at many moments throughout the day.

The emergence of agency in language greatly increased the importance of the shadow audience. When mentally rehearsing any turn in a conversation, we use an imagined shadow audience to infer how other people will understand the turn. When language has agency, the shadow audience makes us responsible for our acts (intended or done), and can infer our resulting status. With every turn in the game – even imagined turns – we infer our own status, as seen by a shadow audience. The result is our fluctuating self-esteem, as we think other people see us.

This rubicon of the mind is captured in the Genesis story of the Garden of Eden. When Adam and Eve were given knowledge of Good and Evil, it was not knowledge of some absolute good and evil; it was good and evil relative to the norms of their social group, used to compute social status in the group. The agency of language makes it possible to attribute acts, together with their good and evil, to people. It makes blame and praise possible. In the Genesis myth, God stands for the all-seeing shadow audience of the mind – knowing all we think or do, and judging our status.

When language started to attribute agency to people, was there some related change in our brains, that gave us genuine agency, a free will not granted to animals? Real free will seems to run up against a logical problem of infinite regress: if you freely decide to act, how do you decide to decide? This has been the subject of philosophical debate, and I shall not discuss it further.

**Stage 6: Spoken Language and Increased Complexity**

As gestures, gaze and the other tools for communication have limited bandwidth and expressive power, the intention game might then have evolved in any of several ways to increase its complexity and make it a more stringent test of brains. Sexual selection always has an element of arbitrariness, reinforced by positive feedback. Human evolution chose the voice channel to increase the bandwidth of language – possibly because of its superior broadcast capability, and its continued usefulness after dark – but to some extent, it may have been an arbitrary branch in our evolution, locked into the human genotype by sexual selection.

In this stage, voice became the pre-eminent interaction channel, while other channels were not abandoned. This allowed yet more complex information to be conveyed within each turn of a conversation – and conveyed more quickly.

One difference between the voice channel and gesture is that gestures can stand in an analogue relation to spatial, consciously perceived reality, whereas voice cannot. Vocal representations of meaning must be encoded in some arbitrary manner, making them harder to learn; but still within the fast learning capability of animals and humans. The arbitrariness of voice encodings opens up a greater possible range of meanings that can be encoded – allowing the intention/status game to become yet more complex.

In this way, very complex meanings can be conveyed in one turn of the intention game, and the game becomes a yet more stringent test of intention reading. The scope of possible meanings increases, and larger brains are required to play the game. Sexual selection for intention reading is now approaching a stable point, where the handicap of a large expensive brain is in balance with the benefits of greater reproductive fitness. We have almost reached modern *homo sapiens*. The intention/status game has become a large set of group-specific language games [Wittgenstein 1953].

### Summary of the Stages of the Intention Game

I have outlined six possible stages of the intention/status game, in a way that might imply that they occurred serially in that order; but we cannot be confident about the ordering and possible overlapping of the different stages.

A more realistic picture is that certain capabilities have grown from zero to their modern human values, each in a wedge-like manner, and that we only know a few constraints on the timings at which the different wedges started. This picture is shown in figure 6.

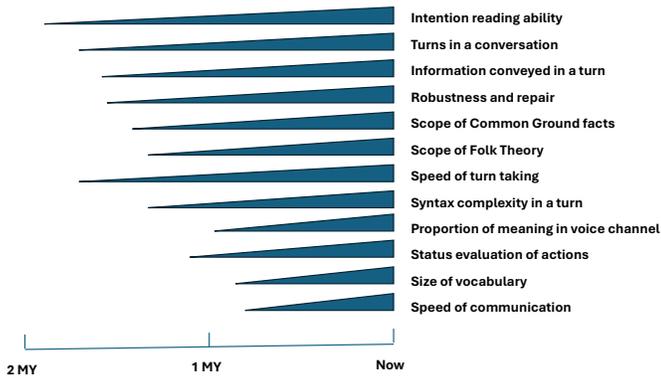

*Figure 6: Possible onset of language capabilities*

(add shadow audience) The timescales, and the starting order of the wedges, are merely illustrative. The growth of the wedges is not expected to be as linear as in the diagram. Readers are invited to vary the picture.

Figure 6 can be likened to the time course of an explosion of a complex mixture of chemicals. We know that there is strong positive feedback, as the heat released by burning one chemical causes another chemical to ignite; this leads to the explosion. But the precise ordering of the burning of the different ingredients cannot be easily predicted. In the case of language, the positive feedback is driven by sexual selection.

This describes the macro-evolution of the intention/status game as a human capability, over millions of years. The detailed forms of the game have varied between hominin groups and over shorter historic times, through a micro-evolution process (the evolution of constructions, not of people) which will be described in section 9.

### Music and Dance

While conversation was evolving though the stages sketched above, there was a parallel evolution of another important part of the intention game. One person can place a new fact in the common ground by starting some rhythmic activity – beating a stick on a tree, or moving their whole body. Another person can respond to this by taking up the same rhythm, in some similar or related activity. In this way, they can show off sharing an intention – they can jam together.

The benefit of a rhythm over one-off ostensive communication acts, is that once you find the rhythm, you can take your own time in various responsive actions tied to the rhythm. You can try out and convey 'messages' at leisure, rather than be constrained by the 200 millisecond turn-taking time of conversation [Levinson & Torreira 2015]. You can share with someone else, at a more emotional than intellectual level.

So in parallel with conversation, music and dance have evolved as another test of shared intentionality; they are not as intellectually competitive as language, but they are still a part of the strong sexual selection forces which have made us human, an important part of our psyche, and a driver of our emotions. They can take place in dyads, in small groups, or social institutions like bands. In institutional settings, they can be a component of religious and other rituals.

There is an emotional dimension to any rhythmic activity. When taking part in any such activity, you may experience a diminished sense of agency – that the rhythm is doing it to you, rather than your own self doing it as agent. This partial release from agency may have positive emotional effects.

### The arbitrariness of sexual selection

In reading this account of the evolution of language, you may feel that some aspects have been included *ad hoc*, because they are parts of modern language, and not because they follow inevitably from the sexual selection hypothesis. That may be true, and it cannot be avoided - because to some extent, sexual selection is arbitrary.

This can be understood by an analogy. Sexual selection is the origin of the gaudy plumage of tropical birds; they display plumage to get a mate. This account can tell us a certain amount about the plumage of any species – that it is vividly coloured, unlike a more drab camouflage, easy to show off and to size up, and differing from the plumage of any other species. But it cannot tell us the detail of the plumage of one species – the precise colours and shapes. It cannot do so because the results of sexual selection are to some extent arbitrary; some initially small fluctuation has been amplified, in random directions, by the positive feedback of sexual selection, until it is locked into the species genotype and phenotype.

There is no detailed scientific account of the plumage in figure 7 below.

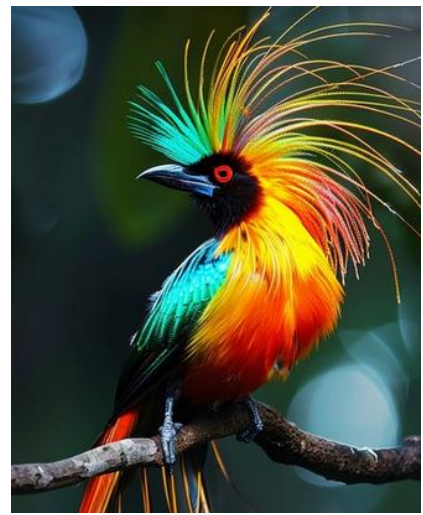

*Figure 7: plumage of a tropical bird. There is no scientific account of the details.*

In the same way, the sexual selection account of language can tell us certain things about it – that it involves competitive displays of intention reading, which over evolutionary time became more prodigiously powerful and fast – but we cannot say exactly in what ways the game has become so powerful and fast. Some details of the intention game simply are how we observe them today, and there is no more detailed evolutionary account of them. That is just how it is. It is a limitation of science, if we have no fossil record to trace the course of evolution.

## 9. Language Change and Diversity

It is not possible to understand modern language (or to understand its precursors in the intention game) without understanding processes of historic language change [Dediu et al 2013; Chater & Christiansen 2010; Christiansen & Chater 2016].

The clearest understanding of language change comes from the following analogy with biological evolution:

- A speaking community or group is like an ecology.
- Each construction in a language is like a species in that ecology
- A language is fully defined by the set of its constructions; all its grammar is embodied in the constructions.
- One person's mental representation of a construction is like an individual of the construction species
- The feature structure graph for a construction is like the DNA of the species
- Each construction reproduces itself by being used by people, and learned by other people
- The learning guarantee ensures that the DNA of a construction (its feature structure) is preserved precisely as it reproduces, with only occasional errors. The learning guarantee is like DNA replication.
- Each language ecology can support a large (but not unlimited) number of constructions.
- Construction species compete for pragmatic and semantic niches within the ecology
- New construction species arise all the time, for instance by social or geographic or technical changes in the group, which create new meaning niches.
- There are several different selection pressures on a construction species, including:
    o Competition with other constructions in its niche, or in overlapping niches
    o Brevity of expression
    o Early resolution of ambiguities
    o Ease of learning
- These are strong selection pressures; any construction species may thrive, or change, or split, or become extinct, within a few human generations.

- This has resulted in the huge diversity of languages, with profound differences between them, but with a small number of universal features
- Language universals tell us more about the process of language change, than they do about the human brain.

This evolutionary model applies both to modern language, and to the earlier stages of the intention game – when feature structures, the communication guarantee, and the learning guarantee were all essential components of the game.

This model of language change was applied to syntax and semantics by Worden [2002]. I shall use an example from there to illustrate the importance of 'hard' ambiguities as a selection pressure on constructions.

[Greenberg 1963, Hawkins 1994] and others have found universals of language structure, some of which hold with high statistical reliability across nearly all known languages [Evans and Levinson 2013]. Typical of these is Greenberg's Universal Number 2:

> *In languages with prepositions, the genitive almost always follows the governing noun, while in languages with postpositions it almost always precedes.*

In a preposition language like English, for instance, the phrase '*the lid of the box on the table*' refers to some kind of lid. However, there is a structural ambiguity, in that the phrase can be parsed in two different ways:

1. ((the lid of the box) on the table)
2. (the lid of the (box on the table))

In a unification-based construction model of grammar, these correspond to two different orders of unification. Both readings mean some kind of lid; the first reading asserts that the lid is on the table (the rest of the box might be somewhere else), while the second reading, it is the box (including the lid) which is on the table.

Because both readings define a lid and its location, this is called a 'soft' ambiguity; you can carry on understanding the whole utterance about a lid, and come back later to the issue of the box and the table, as finer details of the lid. In Japanese, which uses postpositions and obeys universal number 2, it is also a soft ambiguity. But in any language which does not obey universal number 2, it would be a so-called 'hard' ambiguity; you do not even know if the phrase refers to a lid. In the second parse, it is a table – so all you know is that some kind of thing is involved.

Violations of Greenberg's universal number 2 make some utterances very hard to understand – specifically, they make them hard to understand early, to start planning your own quick response. This provides a strong selection pressure on many constructions in a language, to 'line them all up', like atoms in a magnetic domain, in one direction (prepositions) or the other (postpositions).

This evolutionary model goes a long way to account for the peculiar mixture of regularity and irregularity which we see in all the world's languages [Worden 2002]. In the model, the constructions of a language are units of information which reproduce by cultural transmission. In a fairly precise sense, they are memes [Dawkins 1976].

The evolution of constructions applies not only to modern language. It applied to the previous stages of the intention game, when the game was more about pragmatically using facts already in the common ground, than it was about using facts introduced by the latest utterance. In these earlier stages of the game, there was a strong selection pressure on constructions to avoid hard ambiguities; there was pressure on each construction to allow early resolution of its turn type, to allow the planning of a prompt response.

These selection pressures have continued in modern language – forcing each pragmatic construction to allow early resolution of its broad speech act type. These pressures may have led to pragmatic universals, analogous to the Greenberg-Hawkins universals. Such universals have not been discovered yet, because the necessary cross-cultural studies of pragmatics have not been done.

Amongst the processes of evolution of constructions, there may be an evolution from multi-turn pragmatic constructions to single-turn syntactic constructions, which allow faster communication.

Language change may be like a river – a river whose tributaries split downstream, as well as converge. Looking for a neat distinction between semantics and pragmatics is like trying to carve the river into rectangles. A few feet downstream, or human generations later, the rectangles have all merged with each other. Pragmatics has merged into syntax and semantics.

The nature of language change gives strong reasons to doubt the existence of any pragmatic/semantic divide. Almost any construction in a language contains both semantic and pragmatic information; there is a continuum from mainly pragmatic constructions, to mainly semantic. Over a few human generations, any construction may drift from one side to the other of this continuum. It may appear at different places in the order of unification of an utterance, both in production and comprehension. In the presence of all this intermixing, pragmatics and semantics are inextricably intertwined in processing; a clean pragmatic/semantic boundary is not possible.

In section 8, I described how each language is embedded in its own folk theory of the world – a set of assumptions about how the world works (including a folk physics, folk medicine, folk psychology, and so on) which are used in language understanding. The folk theory of the world is also subject to historic change over a few generations. But the replication of parts of the folk theory is not subject to the learning guarantee (that language constructions replicate precisely over the generations, like DNA) so the rate of change of the folk theory of the world may be faster than the rate of change of a language. Today's English folk theory of the world differs from the Shakespearian folk theory of the world, more than modern English differs from Shakespearian English.

Language is the result of two overlapping evolutionary processes: human sexual selection for proficiency in fast mind-reading and unifying constructions; and the selection of language constructions, as described in this section. The second evolutionary process happens much faster than the first – a language construction evolves in a few human generations, as opposed to millions of years (the time required for large changes in the human brain). This difference in speeds is about a factor 1000. Many of the features we observe in modern language, such as language universals, have been shaped much more by the second, faster process, than by human evolution. Language may be telling us more about human culture than about the human brain.

## 10. Construction Grammars Extend to Handle Pragmatics

The study of language pragmatics (how we carry on conversations) has lagged behind the study of syntax and semantics. Pragmatics has been regarded as a poor relation to those two more 'rigorous', intellectually clean topics. It is widely held that syntax is an autonomous module of the mind, and can be studied on its own. This has contributed to the common view of pragmatics – that syntax and semantics each do their job, and then toss over the wall a logical meaning representation, for pragmatics to 'enrich' in some messy, context-dependent way.

The evolutionary insight that language began with context – with pragmatic skills rooted in a common ground of understanding – shows that that research agenda no longer makes sense. Pragmatics started when there were no syntactic or semantic modules, to toss a logical meaning structure over the wall to it. A new approach is required.

The new approach has been hiding in plain sight for several years. It is cognitive linguistics; which, as in earlier sections of the paper, already does a good job on syntax, semantics and language learning. Only a modest extension of construction grammars is required: the unifications of construction grammar must be unifications with facts in the common ground, placing new facts in the common ground; many constructions are pragmatic constructions, feature structures with their 'phonetic' poles containing facts in the common ground. In this picture, all the unifications of language are a form of Bayesian pattern matching within the common ground; when two people converse, they use the same stock of language constructions, and unify them so as to keep their two versions of the common ground in step with each other. This is successful communication.

This picture of pragmatics differs from conventional pragmatics (either neo-Gricean, or relevance-theoretic pragmatics) in several ways.

- The inferences done by a speaker (by unification) are as important as the inferences done by a listener (which have previously been the main topic of pragmatics);
- Pragmatics, semantics and syntax are a seamless whole, done by overlapping unification operations.
- As soon as any pragmatic construction has been learned, slow general-purpose pragmatic inference is replaced by fast unification.
- Many learned constructions have a mind-reading component, giving us a 'Fast Theory of Mind' capability (which is now a central part of human nature).

Before describing particular pragmatic phenomena, and how extended cognitive linguistics accounts for them, I first define the framework of extended cognitive linguistics, and introduce a semi-formal notation for its constructions and operations. In setting up this notation, this is an extended section of the paper.

### Syntax and Semantics in Cognitive Linguistics

In conventional cognitive linguistics, each construction is represented by a feature structure (a directed acyclic graph with information slots on its nodes) with two main branches, or poles, which are called the phonetic pole and the semantic pole. Envisage the construction as a feature structure with the phonetic pole on the left, and the semantic pole on the right.

The phonetic pole of a simple noun construction contains only the sound of the noun. The semantic pole contains the meaning of the noun – a feature structure representation of some thing, with slots for its attributes.

The phonetic pole of a simple verb construction contains not only the sound of the verb, but also some semantic sub-branches, representing the things involved in the verb action, in different semantic roles (such as agent or patient). Every verb construction includes at least an agent thing, represented as part of its phonetic pole. Its semantic pole brings together the things and the action in the full meaning of the verb scene; the semantic pole is a pure meaning, with no word sounds.

In language understanding, each construction is unified from left to right – starting with constructions for the nouns (whose left-hand sides are sound-only), and ending with other parts of speech (such as verbs) which need to have some noun meanings on their left-hand 'phonetic' pole in order to match. The result of this sequence of unifications contains a complete semantic meaning structure at its right-hand (semantic) end.

For language production, unification goes in the reverse order – each construction unifying from right to left (from semantic pole to phonetic pole), and in nouns-last order. Each unification is done by matching the semantic pole, and adding the phonetic pole. The result of all the unifications, when the all the meaning parts of phonetic poles have been matched, contains some word sounds towards its left-hand end, which are to be spoken.

The result of these paired complementary processes for production and understanding is the communication guarantee, covering syntax and semantics – a mathematical guarantee that the listener can successfully re-create the speaker's intended meaning, as described in the appendix.

These paragraphs are a very terse description of some complex mathematical operations. A more complete description, with online working examples (feature structures that can be inspected) and the proof that it all works, can be found at [Worden 2022a].

To extend this successful model of language from syntax and semantics, to address pragmatics as well, one new ingredient is needed: to take proper account of the common ground shared by speaker and listener; to make a model of how the common ground is represented in their minds, and is used in language.

### The Common Ground

The term **common ground** was introduced by [Stalnaker 1978], and the concept has been extensively investigated by him and others [Clark 1996; Tomasello 2009,2014; Stalnaker 2002, 2014]. The common ground is a set of facts and other knowledge which are shared between a speaker and a listener (interlocutors) in a conversation.

The common ground is central to this theory of language. This section describes what is in the common ground, and how it is structured.

The common ground consists of the following:

1. Facts which are obvious to the interlocutors from their sense data, represented as feature structures.
2. Facts in their shared history, represented as feature structures.
3. World knowledge (including knowledge of their social milieu) which the interlocutors share, represented as facts and associative rules. These rules are represented by feature structures with no word sounds, only meaning structures.
4. Facts which can be derived from (1) and (2), using (3) – both directly and indirectly, by recursive

multiple unifications of the rules. These are called **latent facts**.
5. Shared linguistic knowledge: a set of constructions A⇔B (where A is the phonetic pole, and B is the semantic pole) in which pole A contains one or more word sounds (and multi-modal gestures), and possibly other meaning structures. B contains only meaning structures.

Facts which are not latent, but are explicitly represented and stored as feature structures in the common ground, are called explicit facts.

It is assumed that both interlocutors initially have the same internal representation of the common ground – the same set of feature structures - and their internal representations stay in step as they converse, by mechanisms described below. As the foundation for this, it is assumed that the interlocutors have the same set of world knowledge rules (3) that they have learned, and they have learned the same set of linguistic knowledge rules (5). They have the same set of immediate facts (1), and the same facts in their shared history (2).

Because of the recursive nature of (5), the common ground contains an unbounded set of facts – in principle, an infinite number of different latent facts can be derived by unification. This raises the uncomfortable prospect that inferences using the common ground will require infinite computation, which is computationally infeasible.

To avoid the threat of infinite computation, every fact in the common ground is assigned a measure of its **prominence** (denoted by Q), which is a number between 0 and 1. Highly prominent facts, such as things or events which are currently visible to both interlocutors, have prominence near 1.0. Less prominent facts have prominence near zero. The prominence of a fact relates to the probability that it will be used in a linguistic inference.

If fact A is in the common ground and has prominence $Q_A$, and a latent fact B can be inferred from fact A by using a rule A => B, then the prominence $Q_B$ of B is less than $Q_A$ by some factor, whose precise details are to be defined; for instance, we might assume that $Q_B = 0.5* Q_A*P(B|A)$, so that B is always less prominent than A. This applies recursively to multiple applications of the rules, giving decreasingly prominent latent facts.

The concept of prominence is what makes the common ground computable with finite resources; prominence makes the common ground computable, and the common ground makes language possible. Prominence evolved from Bayesian probability, which is commonly used in animal brains [Rao et al 2002; Friston 2010]. Prominence is a measure of the prior probability that some fact will be used in interpreting the next exchange between interlocutors. Prominence is an essential property of facts in the common ground, and it has evolved in human brains to make language possible.

Prominence behaves mathematically in some ways like a Bayesian probability. For instance, if there are two facts A and B in the common ground, with prominence $Q_A$ and $Q_B$, and there is a rule [A & B] ⇔ [C], then the latent fact C has prominence $Q_C$ equal to $0.5*Q_A*Q_B*P(C|A \& B)$.

In the Free Energy Principle (FEP) [Friston 2010], Free Energy in the brain is the negative logarithm of a Bayesian probability. So the negative logarithm of the prominence (a positive number) can be regarded as a kind of free energy, specialized to language and the common ground. In this way, it may be possible to apply the computational techniques of the FEP (such as Active Inference [Parr, Pezzulo & Friston 2022]) to language inferences in the brain. Active inference is the minimization of a Variational Free Energy (VFE) for perception, or of an Expected Free Energy (EFE) for planning. If we define Prominence Free Energy (PFE) as the negative log of the prominence, then linguistic inference in the common ground is minimization of a PFE.

The facts in the common ground can be envisaged as a cloud, with a few prominent facts at the centre of the cloud, and progressively less prominent facts as you go further away from the centre.

Any inference procedure which uses facts in the common ground is implemented depth-first and cloud-centre-first, to use facts in order of decreasing prominence, with some termination criterion which will be illustrated below. This means that common ground inferences can be made with finite computing resources.

Prominence may be related to the concept of relevance in relevance-theoretic pragmatics. The concept of relevance has not been quantitatively defined, but it plays an important role in the rules for relevance-based pragmatics; you make inferences until you find a conclusion which is sufficiently relevant; then you can stop. This seems to have a link to inference in order of prominence, but I have not yet stated the link precisely.

Facts and rules in the common ground can be rapidly retrieved based on their content, by **associative retrieval**. For instance, given some construction A⇔B, it is possible to rapidly and in parallel retrieve all facts in the common ground with prominence greater than $Q_0$ which match A, to unify those facts and compute all the resulting B, in parallel (or to do the same with A and B reversed). Similarly, given some fact F, it is possible to retrieve and unify in parallel all rules A ⇔ B in which A matches F, or in which B matches F. The maximum amount of parallelism may be limited in animals; but in humans, greater parallelism and speed may have evolved competitively to support fast language, enlarging our brains.

Unification is a fast, parallel, pre-conscious operation. Evidence from language comprehension (which, in cognitive grammars, is done by unification) implies that several unifications can be done in parallel, and that for chained serial unifications, the chains can extend at a rate of about 5 links per second. This includes the time required for associative retrieval of a few appropriate constructions, out of many thousands of constructions (a person has at least one language construction for every word she knows; and possibly a greater number of world knowledge rules).

Because the prominence of a latent fact contains a factor less than 1 for each rule that has to be unified to infer it from a explicit facts, the negative log of its prominence is an approximate measure of the **cognitive effort** required to make it prominent; it is approximately the number of serial unifications required. The number of serial unifications is proportional to the elapsed time required to infer the fact. It is also, as above, the Prominence Free Energy (PFE) of the fact. This may also relate to the principles of relevance-theoretic pragmatics.

### Unification in Pragmatics

To describe pragmatic devices, the unification procedure for both speaker and listener remains as it was for the application of construction grammars to syntax and semantics, with the following changes:

- When someone utters a word, the sound of the word becomes a fact in the common ground, with high prominence, and with a time label.
- The only unification which occurs, for speaker or listener, is unification of facts in the common ground. This is done in order of their decreasing prominence, with some termination criterion.
- Both sides of a construction need to be unified with facts in the common ground, and the result of any unification is placed in the common ground as a new fact with high prominence.

These points are illustrated below with examples, to show how they work – first in some simple cases of deixis (resolving noun reference) and then for other pragmatic devices.

### Notation for Constructions

The full notation for facts, constructions, and rules uses the tree notation for feature structures, requiring diagrams (as for instance, seen at [Worden 2022a]). As a simpler textual notation, I use the following conventions:

- A (non-linguistic) fact is denoted by upper case words in brackets; for instance [MALE & YOUNG] denotes a young male person; [ANNA] denotes a named individual; [JOE SITS] denotes a state. Nesting is possible, as in [ANNA WANTS [ANNA SITS]], representing a Theory of Mind fact. Logical operators are sometimes stated, sometimes assumed. Time ordering of facts is denoted by commas, if their time order is constrained. Slot values can be denoted by GENDER:FEMALE if the slot name is not obvious.
- Variables in facts and rules are denoted by single upper-case letters like X. If unification fixes a variable in one part of a fact or rule or construction, the variable is fixed to the same value everywhere in the fact or rule or construction.
- The probability that a fact is correct is denoted by P, and its prominence in the common ground by Q; for instance, $P = 0.8$, $Q = 0.25$. P and Q are independent. P is the probability that a fact is true; Q relates to the probability that it will be used in a language inference.
- A pure sound fact (a word spoken, heard and so in the common ground) is denoted by a lower-case word in brackets, such as ['fred'].
- There can be compound facts containing both word sounds and meanings.
- A rule of world knowledge is written as for example in $[EGG]_A \Leftrightarrow [FOOD]_B$, and $[X: FOOD \& COOK\ X]_A \Leftrightarrow [EDIBLE\ X]_B$ with the two sides of the rule defined by indexes A and B. Conditional probabilities of rules are defined as in $P(B|A) = 0.7$ or $P(A|B) = 0.2$. If a probability is not stated, it is zero.
- The usual use of a rule is to unify its A side with some fact or facts in the common ground (possibly fixing its variables), adding the resulting B side to the common ground; or to do the same with A and B sides exchanged.
- When a rule is applied, the probability of the resulting fact is the probability of the input fact multiplied by the conditional probability of the rule.
- There is some other way (not yet fully defined) to define a prominence for the result of applying a rule
- Facts and rules are given labels, such as F1 for facts, Wo1 for world knowledge, and La1 for language rules (constructions).

Any facts in the common ground are known by both interlocutors. This is encoded as two 'Theory of Mind' rules:

Wo1: $[X]_A \Leftrightarrow [SPEAKER\ KNOWS\ [X]]_B$ $(P(B|A) = 1.0)$

Wo2: $[X]_A \Leftrightarrow [LISTENER\ KNOWS\ [X]]_B$ $(P(B|A) = 1.0)$

Here, X denotes any fact in the common ground. These rules can be applied repeatedly to generate an infinite number of nested Theory of Mind facts, of ever-decreasing prominence. The decreasing prominence effectively solves the problem of infinite regress, for theory of mind facts; inferences are made in order of decreasing prominence, and the sequence of inferences always terminates.

If the SPEAKER knows some fact [X], X may be in the common ground, but it need not be; the probability for right-to-left application of these theory of mind rules is less than 1, possibly zero.

A simple noun construction is

La1: ['bert']$_A$ ⇔ [BERT]$_B$     $P(B|A) = 1.0$

This construction (which has label La1) says that if the sound 'bert' is in the common ground (i.e. if the word has been spoken), and there is some person X in the common ground, which can unify with the person with identity BERT (i.e. the properties of person X are consistent with the known properties of BERT), then the unification result [BERT U X] is added to the common ground – so any existing properties of the person X are attributed to BERT. If the properties of person X conflict with the properties of BERT, then the unification fails, and the rule has no effect. Persons X in the common ground are unified with BERT in the order of decreasing prominence, until some unification succeeds. Then the process stops.

In the pictures of language use which follow, the textual notation for facts and constructions is further simplified:

- Facts are written in single rectangular boxes
- Constructions are written in joined pairs of boxes, with the phonetic pole on the left (shaded in yellow) and the semantic pole on the right (unshaded).
- Non-linguistic rules are written a joined boxes, with both halves white

Language production proceeds by matching the semantic pole, and adding the phonetic pole to the speaker's common ground (when the speaker says the word, it comes part of the shared common ground); while comprehension proceeds by matching the phonetic pole, adding the semantic pole to the listener's common ground.

Constructions like La1, and more complex rules, can be viewed as feature structure diagrams (directed acyclic graphs) in [Worden 2022a].

The phonetic pole of a construction can contain gestures, facial expressions, and indicators of vocal tone, as well as word sounds. There are often time-ordering constraints between these parts, restricting the facts in the common ground the construction can unify with.

When some entity in the common ground is picked out by speaking a word, or by pointing at it, or by gaze, that entity is promoted to high prominence $Q = 1.0$, and any entity which previously had prominence $Q = 1.0$ is demoted to a lower prominence such as 0.8, and so on for all entity facts. Similarly for event facts in the common ground.

[Stalnaker 2014] defines a **subordinate context** as a subset of the common ground which is applicable in a particular context, such as [JOE THINKS] or [BEDTIME] or [RESTAURANT]. In this model, a subordinate context has the same world knowledge rules as the main context (for instance, it has a set of rules equivalent to Schank's [1974] 'Restaurant script', concerning WAITER, MENU, MAIN_COURSE and so on; all these rules depend on the predicate RESTAURANT, combined by logical '&' with other conditions). So the prominence of all latent facts derived from these rules is multiplied by the prominence of some root fact, such as RESTAURANT. Thus after any mention of a restaurant, the prominence of RESTAURANT, and of the whole set of latent facts which depend on it, is immediately increased; when a subordinate context comes to the foreground, all its facts become more prominent.

During any utterance, the speaker's common ground is continually altered, as each word is added and as it triggers language rules. New explicit facts are added, and the prominence of facts is changed. As the listener understands the utterance, his version of the common ground undergoes a series of alterations with successive unifications. The important fact is that these two series of unifications keep the speaker's and the listener's version of the common ground in step, so that conversational repair is rarely necessary.

**Speakers' and Listeners' Procedures**

We can now define the joint model of inference for speakers and listeners. Here I give a general description, including features which are sometimes needed for a treatment of ambiguities, and vocabulary shortcomings.

The steps done by a (female) speaker are:

> S1: The speaker wishes to express a meaning M, which is in her mind, but not yet in the common ground, or is in the common ground with low prominence.
>
> S2: She retrieves all constructions R whose semantic pole matches with M, associatively by matching the meaning M.
>
> S3: She applies each language rule by unification of facts in the common ground (in order of decreasing prominence) with the meaning M, to make a specific version of the phonetic pole A of the construction. This contains word sounds W and (for words other than nouns) other meaning parts N, O, P, …
>
> S4: (dealing with ambiguity) The speaker retrieves any other constructions R', whose phonetic pole A includes the word sounds W of rule R (these are the ambiguous senses of the words W), associatively by the word sound W.
>
> S5: (dealing with ambiguity) for each such construction R', the speaker unifies its semantic pole B with facts in the common ground, in order of decreasing prominence, until one of them matches, giving meaning M'.

S6: (dealing with ambiguity) if M' is equal to M, and the unified facts are the most prominent of any rule R, R'…, the rule R is an appropriate one to apply. The speaker retains the meaning parts N, O, P from the sound-meaning side of the rule (for nouns, there are no such meaning parts N).

S7: She puts the meaning M into her own version of the common ground with prominence 1.0, or promotes it to prominence 1.0, demoting the prominence of all other facts by some factor (e.g. 0.8).

S8: for each meaning part N created in step S6, the speaker recursively applies steps S2 to S7, with N replacing M, until all meaning parts N have been matched by language rules.

S9: she speaks the sequence of words sounds – putting the word sound facts into the common ground, for both herself and (through hearing the words) for the listener. The order of words is defined (if it is defined) by time-ordering constraints in the language rules – not by the order of unification of the rules.

The (male) listener's inference process is as follows:

L1: For each heard word W, he retrieves all constructions R, R' etc. which contain the word sound in their sound-meaning side, and unifies them with facts in the common ground. There is a priority order for these unifications. Words with fewer meanings in the phonetic pole are unified first. So the first constructions to unify are those for nouns, which (like the example rule for 'fred') have no meaning part in their phonetic pole. Rules for other words like verbs have some meaning parts in the phonetic pole

L2: a construction R may unify with several facts in the common ground. For each R, the facts are unified in order of decreasing prominence, until one is matched. Then the process stops.

L3: For the construction R which unified with the fact with the greatest prominence, the resulting fact (in the semantic pole) is added or promoted in the listener's version of the common ground to prominence 1.0. Other facts are demoted (their prominence multiplied by a constant factor less than 1.0).

L4: when all noun constructions have been unified (giving a number of high-prominence entities in the listener's common ground), the listener applies steps L1 and L2 to other heard words W, whose phonetic pole contains meanings as well as word sounds (in the priority order of step L1) These meaning parts can be unified with recently promoted facts of high prominence (e.g. meanings of nouns), producing new meanings which are added to the listener's common ground with high prominence.

L5: The listener proceeds recursively until all word sounds have been matched.

The effect of these two inference procedures is that the speaker unifies constructions in a top-down, nouns-last order, while the listener unifies the same constructions in the reverse, nouns-first order. The speaker matches the semantic pole of each construction, adding the phonetic pole to the common ground. The listener matches the phonetic poles of the same constructions, adding the semantic pole to the common ground. At the end of this process, provided their two versions of the common ground were initially in step, the two versions will remain in step. The same set of meanings will be added to the common ground for the listener as for the speaker, with the speaker's intended meaning M having the highest prominence. This is the communication guarantee.

The key to the communication guarantee is that the speaker and the listener use the same constructions – the speaker right-to-left, and the listener left-to-right. The listener's inference process (words => meaning) mirrors and reverses the speaker's inference process (meaning => words).

[Clark 1996] has emphasized that conversation is a joint activity of the speaker and the listener. This is a model of their joint inference to underpin the joint activity.

The recursion enables this joint model of speaker's and listener's inferences to handle production and comprehension of utterances with complex syntax, and with ambiguities, but it makes it harder to see how the communication guarantee works.

To illustrate how the guarantee works, the next section describes how it works in simple non-recursive examples (one-word utterances), with little syntax – but with possible ambiguities.

## 11. First Pragmatic Phenomena: Deixis and Early Syntax

To illustrate the communication guarantee, I shall use examples of **reference resolution** – where language is used only to pick out some entity in the common ground and increase its prominence. This entity can be described by a single noun, or pronoun, or noun phrase, or noun clause (the last two cases are not considered here). The pragmatic problem is: how can the speaker describe some object, and be sure that the listener will understand the same object?

While in general the common ground can contain both objects and events, in these examples only objects are shown.

In the first example, before the utterance, the common ground contains the following labelled facts and rules (the facts are certain, with probability P = 1.0):

(CG1) Immediately perceptible facts:

F1: [JOE]              Q = 1.0

F2: [ANNA]             Q = 0.8

F3: [FRED]             Q = 0.6

F4: [X: TREE]          Q = 0.4

(CG3) World knowledge:

Wo1: [ANNA & PERSON & FEMALE]

Wo2: [JOE & PERSON & MALE]

Wo3: [FRED & PERSON & MALE]

(CG5) Linguistic knowledge: (constructions)

La1: ['fred']$_A$ ⇔ [FRED]$_B$         $P(B|A) = 1.0$

La2: ['joe']$_A$ ⇔ [JOE]$_B$           $P(B|A) = 1.0$

La3: ['he']$_A$ ⇔ [MALE]$_B$           $P(B|A) = 1.0$

La4: ['she']$_A$ ⇔ [FEMALE]$_B$        $P(B|A) = 1.0$

Suppose that the speaker wishes to pick out the individual FRED, who is known to both interlocutors.

Using the step labels (S1).. and (L1).. of the joint inference model of the previous section, the speaker's and listener's inference processes go as follows:

(S1) The speaker's intended meaning is [FRED].

(S2) because the speaker's world knowledge includes the fact Wo3: [FRED & PERSON & MALE], the applicable linguistic rules are those whose meaning-only sides match (can unify with) FRED, or with PERSON, or with MALE. So the applicable words are 'fred' and 'he'. These are found by rapid associative retrieval of the constructions by their semantic poles, from a large vocabulary of constructions.

(S3) For each of these constructions, she tries to unify it with entities in the common ground, in order of decreasing prominence until a match is found:

- The construction La3 for 'he' matches with JOE, and the process stops.
- The construction La1 for 'fred' does not match with JOE or ANNA, but it matches with FRED, and the process stops.

(S4) in the speaker's set of constructions, there are no other constructions with phonetic pole containing sounds 'fred' or 'he', (e.g. she knows no other senses of the word 'fred' or 'he') so this step retrieves no other rules, and step (S5) is not needed.

(S6) for the word 'he' the expected meaning is JOE, which does not match the speaker's intended meaning FRED, so the word 'he' cannot be used. For the word 'fred', the expected meaning is FRED, which is the speaker's intended meaning, so the word can be used.

(S7) The speaker promotes fact F2: [FRED] to prominence 1.0 in her version of the common ground, and demotes other facts.

(S8) in this example there is no recursion, so this step is empty.

(S9) the speaker says the word 'fred', putting a sound fact into the listener's common ground, and into her own common ground with prominence 1.0.

The listener's inference process is as follows:

(L1) There is only one heard word W = 'fred', and there is only one construction La1 with that word sound in its phonetic pole. The construction is retrieved, by associative retrieval using the word sound in the phonetic pole.

(L2) Facts F1..F4 in the common ground are unified with the construction La1 in order of decreasing prominence. The unifications with the facts 'JOE' and 'ANNA' fail because they do not match, but unification with 'FRED' succeeds.

(L3) The listener promotes the fact [FRED] to prominence 1.0, and decreases the prominence of [JOE] and [ANNA].

(L4) There are no other heard words W, so step (L4) is empty.

(L5) The recursive step (L5) is not needed.

At the end of this joint inference process, the fact [FRED] (and the sound 'fred') have been promoted to prominence 1.0 in both speaker's and listener's versions of the common ground, and other facts have been demoted, so their two versions of the common ground remain in step. This is the communication guarantee.

This is illustrated in the following three diagrams. Figure 3 shows the initial state of the common ground:

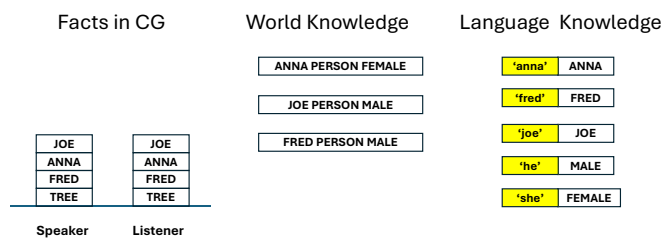

*Figure 3: Initial state of the common ground in example 1*

As noted previously, in these pictures, facts and world knowledge are single boxes, and constructions are double boxes, with the phonetic pole in yellow. In the initial state of the common ground is the same for speaker and listener. Facts in the common ground are shown as a stack, with the most prominent facts at the top of the stack.

Figure 4 shows the speaker's inference process.

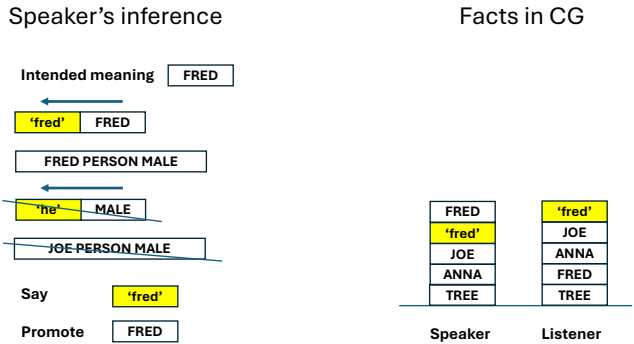

*Figure 4: The speaker's inference process in example 1*

The steps in the speaker's inference process are shown in order from top to bottom of the diagram. The speaker intends to convey the meaning FRED; she uses this to retrieve and apply a word construction. She applies the construction by unification right-to-left (shown by the arrow) leading to a possible utterance 'fred'.

The crossed-out steps on the left of the diagram show a failed line of inference. Since FRED is MALE, a possible word to say is 'he'; but since JOE is currently higher in the stack of facts than FRED (JOE has greater prominence), 'he' would be resolved to JOE, which is not the intended meaning. So the speaker says the word 'fred', and promotes the meaning FRED to the top of the common ground stack. After this, and before the listener's inference, the two common grounds are temporarily out of step.

Figure 5 shows the listener's inference process.

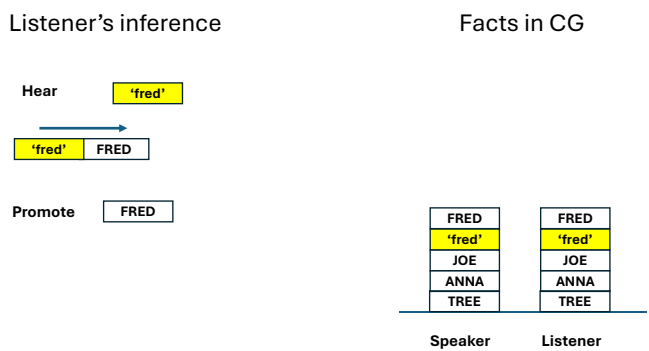

*Figure 5: The listener's inference process in example 2*

The steps in the listener's inference process are again shown in top-to-bottom order. The listener hears the word 'fred' and adds it to the top of his common ground stack. He uses the word to associatively retrieve the construction with 'fred' in its phonetic pole, and applies it by unification left-to-right, as shown by the arrow. This gives the meaning

FRED, which is promoted to the top of the listener's common ground stack – bringing the two common grounds back in step with each other.

You can check that the two versions of the common ground will also stay in step in the following cases:

(1) The speaker wishes to convey the meaning [FRED], and does this by pointing at Fred (to promote FRED to prominence 1.0) and saying 'he'.
(2) The speaker wishes to convey the meaning [JOE] and does it by saying 'he'
(3) The speaker wishes to convey the meaning [ANNA] and does it by saying 'she'.
(4) FRED is not in the common ground, but is in the interlocutors' shared history

I next consider what happens in the case of an ambiguous word such as 'bank', which has two different senses. In this example, there are the following facts and rules in the common ground:

(CG1) Facts immediately perceptible:

F1: [LOCATION: RIVER]          $Q = 1.0$

(CG2) Facts in recent shared history:

F2: [LOCATION: STREET]          $Q = 0.2$

(CG3) World knowledge:

Wo1: $[L: RIVER]_A \Leftrightarrow [L\ HAS\ RIVERBANK]_B$
$P(B|A) = 0.9$   $P(A|B) = 0.8$

Wo2: $[L: STREET]_A \Leftrightarrow [L\ HAS\ MONEYBANK]_B$
$P(B|A) = 0.5$   $P(A|B) = 0.8$

(CG5) Linguistic knowledge (constructions):

La1: $[\ 'bank']_A \Leftrightarrow [RIVERBANK]_B$          $P(B|A) = 0.5$

La2: $['bank']_A \Leftrightarrow [MONEYBANK]_B$          $P(B|A) = 0.5$

Suppose the speaker wishes to express the meaning RIVERBANK. The inference procedure of the speaker is:

(S1) Her intended meaning is RIVERBANK

(S2) she retrieves the first construction La1, by associative retrieval using its semantic pole. That is the only candidate construction with that meaning.

(S3) She attempts to make the unification [RIVERBANK U X ]$_B$ for some X in the common ground, trying facts X in order of decreasing prominence. There is no such X amongst the facts which are immediately perceptible or in the recent shared history, so she searches for some latent fact which is implied by world knowledge. She associatively retrieves the rule Wo1 (because it mentions the entity RIVERBANK) and tries to unify it with facts in the common ground, in order of decreasing prominence. This succeeds with fact F1. The resulting fact has prominence 0.5*1.0*0.9 = 0.45. The search for facts stops then.

(S4) As a possible word is 'bank', she searches associatively for other language rules with the sound 'bank' in their sound-meaning part. She retrieves the language construction La2 for MONEYBANK.

(S5) She tries to unify La2 with facts in the common ground, in order of decreasing prominence. There are no suitable facts (i.e. facts mentioning MONEYBANK) which are immediately perceptible or in the shared history, so she looks for latent facts implied by world knowledge. She associatively searches world knowledge for rules mentioning MONEYBANK, and finds rule Wo2. She tries to unify Wo2 with facts in the common ground, in order of decreasing prominence. There are no suitable facts immediately perceptible, but there is a matching fact F2 (mentioning STREET) in the recent shared history. Unification succeeds and gives a fact with prominence $0.5*0.2*0.9 = 0.09$; so the process stops. This level of prominence is less than that for rule La1, so she knows that the word 'bank' will not be interpreted by the listener as a MONEYBANK.

(S6) The only language rule which gave the intended meaning RIVERBANK is rule La1.

(S7) The speaker adds the meaning RIVERBANK to her common ground with prominence 1.0, and demotes other meanings

(S8) The phonetic pole of La1 gave no other meanings N which need to be expressed, so the recursive step (S8) is not needed.

(S9) She speaks the word 'bank', which puts it into her own common ground and into the listener's common ground.

The two versions of the common ground are temporarily out of step. The listener's inference process is as follows:

(L1) he hears 'bank' and retrieves all constructions which have the word 'bank' in their phonetic pole. These are the constructions La1 and La2.

(L2) For each construction La1 and La2, he tries to unify it with facts in the common ground, in order of decreasing prominence. For each construction, there are no matching facts immediately perceptible, or in the recent shared history, so he looks for latent facts implied by world knowledge rules:

- For construction La1, he associatively retrieves world knowledge Wo1 by its meaning RIVERBANK and matches it with fact F1, giving a fact with prominence $0.5*1.0*0.9 = 0.45$
- For construction La2, he associatively retrieves Wo2 by its meaning MONEYBANK and matches it with fact F2, giving a fact with prominence $0.5*0.2*0.9 = 0.09$

(L3) The fact with greatest prominence is the fact that was created by rule La1, so that fact is added to the listener's common ground with prominence 1.0, and other existing facts are demoted.

(L4) this recursive step is not required.

Once again, the speaker's and the listener's common ground stay in step; and both end up with the most prominent fact RIVERBANK, as the speaker intended. This is the communication guarantee for an ambiguous word 'bank'.

This exchange is summarized in the next three figures. Figure 6 shows the initial state of the common ground.

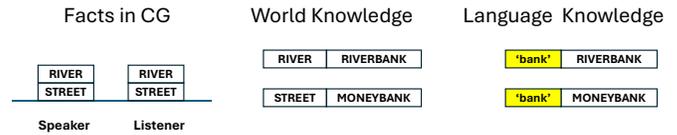

*Figure 6: Initial state of the common ground in example 2*

Here, the facts in the common ground include a current fact (that the location is near a river) and a fact from recent shared history, that the location was in a street. The historic fact has lower prominence; it is lower in the stack.

Figure 7 shows the speaker's inference process.

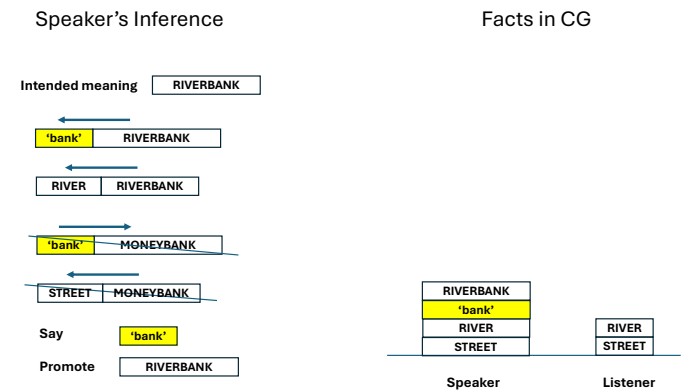

*Figure 7: Speaker's Inference Process in example 2*

Inference steps are again ordered top to bottom in the diagram. The intended meaning is RIVERBANK, so the speaker uses this meaning to associatively retrieve one construction for 'bank', and unify it left-to-right (top arrow) to make a candidate word to speak. She checks this by using the world knowledge rule (again unifying left-to-right, next arrow) to check that the meaning links to the high-prominence fact RIVER in the common ground.

The next two rows on the left show failed inferences, crossed out. She uses the word sound 'bank' to retrieve the second language construction for 'bank', and unifies the construction (this time, right-to-left) to find the other possible meaning in its semantic pole, MONEYBANK. She uses another world knowledge rule to link MONEYBANK to STREET in the common ground; but since STREET has lower prominence than RIVER, this meaning is rejected (the listener will not misunderstand).

Finally the speaker says 'bank', adding the word and the meaning RIVERBANK to her common ground.

Figure 8 shows the listener's inference process.

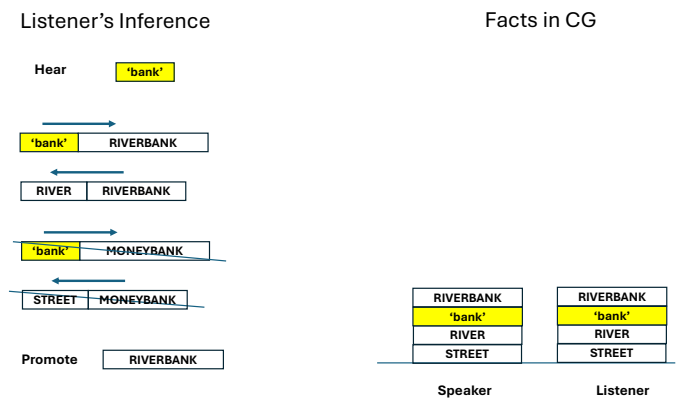

*Figure 8: Listener's Inference Process in example 2*

The listener first hears the word 'bank' and adds this at the top of his common ground stack. He uses the word sound to associatively retrieve two constructions for 'bank', by their phonetic pole. For each one, he unifies it left-to-right to create a meaning, and then uses a world knowledge rule to relate the meaning to facts in the common ground. The first construction gives a result with greater prominence, so the second construction is rejected (crossed-out inference steps). Finally, the listener adds the winning meaning RIVERBANK to the top of his common ground stack. The two common grounds are in step again.

You can check that the communication guarantee also holds in the following cases:

(1) The common ground has current location STREET, and the speaker wishes to convey the meaning MONEYBANK
(2) The common ground has location FIELD (not RIVER), and the speaker wishes to convey the meaning MONEYBANK (because STREET is in the recent history, with higher prominence than RIVER, the word 'bank' can be used).

What happens if the common ground has current location FIELD or STREET, and the speaker wishes to convey the meaning RIVERBANK? There are several possibilities, such as:

a) If there is a river in sight, the speaker can point towards it, so raising its prominence to 1.0; then say the word 'bank'.
b) The speaker can say the word 'river' – thus raising the prominence of RIVER to 1.0, and demoting the prominence of the current location; then say 'bank'

While these examples have mainly used single words, the same principles apply to multi-word phrases whose meanings have been learned as entire sound sequences, such as 'Stone the crows' or 'How do you do?' or 'paragon of virtue' or 'Trojan horse'. These are all single constructions. It is only when words are used productively that the complexities of recursion arise.

With possibility (b) above, using the two-word utterance 'river bank', we see the origin of syntax, in muti-word turns. Another example discusses the earliest stages of syntax.

A similar two-word mechanism works for an intransitive verb, such as 'sleeps'. To understand this fully, you need to use the full diagram notation for constructions; but passing over some details, the construction for 'sleeps' is

[X:PERSON , 'sleeps'] ⇔ [X SLEEPS]

In this construction, the phonetic pole side has a meaning part X:PERSON as well as its sound part 'sleeps'. Note the time ordering constraint in the phonetic pole, implied by the comma. Note how the same variable X appears on both sides of the construction; X acts as a channel to move meanings from one pole to the other. This enables the construction to pull together simple meanings to make a complex meaning (left-to-right, in comprehension), or to break apart a complex meaning into simpler meanings (right-to-left, for production).

To understand the utterance 'joe sleeps', first the construction for 'joe' is retrieved and unified left-to-right, promoting its meaning JOE to high prominence (as in the previous examples). Next, the construction for 'sleeps' is unified left-to-right, with the prominent fact JOE unifying with X:PERSON to give the fact [JOE SLEEPS].

To produce the utterance 'joe sleeps' from the meaning [JOE SLEEPS], the same constructions are unified right-to-left, in reverse order – first the verb rule, then the noun rule. 'sleeps joe' is not allowed because of the time order constraint in the construction for 'sleeps'.

This gives a first glimpse of how the recursive joint inference model gives the communication guarantee, for complex syntax.

The joint inference model applies to fragmentary, ungrammatical utterances. For instance, Joe is eating breakfast, has made some toast, and is ostensibly looking for something. This promotes the theory of mind fact [JOE WANTS [JOE HAS MARMALADE]] to high prominence. When Anna says 'shelf', Joe knows to look for the marmalade the shelf – the most prominent shelf in their common ground. This interchange depends on facts in the common ground such as [JOE MAKES TOAST], [JOE LOOKSFOR X] and on shared world knowledge like [TOAST NEEDS MARMALADE].

[Clark and Wilkes-Gibbs 1986] proposed a model of noun reference resolution as a collaborative joint process between interlocutors. They considered complex noun phrases, and described empirical work on a challenging referring task of identifying Tangram pictures, which typically requires repair and speaker-listener collaboration over several conversational turns. Their model describes this collaboration at a macro level. This section has described a micro-level cognitive model of the required inferences, typically applied to less challenging everyday reference problems, to minimize the need for repair; but this model should also be applicable to more challenging reference problems.

This section has shown how the common ground is used to resolve meanings and ambiguities, in some simple cases; cases for which you may say 'yes, that is obvious'. Introspectively, these examples feel like 'what we do' when speaking and hearing; but they illustrate the mechanisms by which unification in the common ground is used in more complex syntactic and pragmatic phenomena, as described in the next section.

## 12. Other Pragmatic Phenomena

This section describes how the joint model of inference handles a variety of pragmatic phenomena [Huang 2017]. The examples are grouped under some main headings of pragmatics, and there is overlap between the headings.

These descriptions will use an abbreviated form of the notation introduced in earlier sections, for facts, world knowledge, and language rules, to give a flavour of the reasoning used in both speaker inferences and listener inferences. Some detail has been left out. Word sounds are in lower case, and meanings are denoted in upper case; prominence and probabilities are not quantified. Different types of speech act are denoted by labels such as REQUEST for an action, ENQUIRY for some information, ANSWER to an ENQUIRY.

The general tenor of the explanations is as follows:

1. Pragmatic inference is a Bayesian comparison of the relative likelihood of different interpretations, computed using prominence. The prominence of latent facts includes a factor for the probability of every rule used to derive them.
2. This account uses profligate learning of world knowledge facts and rules, and of large numbers of language constructions (in adults, many thousands), for words and multi-word constructions. Accounts of examples tend to be individual and determined by details of the common ground, rather than following from a few broad logical principles.
3. All meanings refer to the common ground, and the search for a meaning is made in order of prominence, using high-prominence facts first, with a termination criterion.
4. Generally, when a pragmatic device is novel, it requires some novel inference – which is sometimes done by unification, but may require more expensive and general processes such as mental simulation, in imagination. After some pragmatic device has been encountered a few times, it is learnt as a construction, and applied quickly by unification.
5. Parsing of complex utterances (by unification in the common ground) is assumed, but not described in detail. Language is typically learned in non-productive early forms (multi-word constructions), with productive forms learned later; that is why discussion of syntax is largely sidelined.

The examples follow.

### Deixis and Reference Resolution

Examples of deixis are given in the previous section. Any fact can be added to the common ground as an explicit fact with initially high prominence. Often new facts are not added, but facts are promoted from being a latent fact with low prominence, to an explicit fact with high prominence. Ambiguity is resolved by looking for the largest prominence; usually this gives a clear-cut answer. Here is one further example of ambiguity resolution:

*Did you order steak?*

The word 'steak' is ambiguous, between RAW_STEAK (in a subordinate context of BUTCHER) and COOKED_STEAK (in a subordinate context of RESTAURANT). Resolution depends on the existing prominence of these sub-contexts; whichever one wins the competition, its prominence is then promoted to near 1.0. The prominences of subordinate contexts depend on the location, recent conversation, shared history, and so on.

### Presupposition

Presuppositions are various ways in which an entity fact or an event fact can be promoted to high prominence. Presupposition often has properties such as surviving negation, and being cancellable or defeasible [Levinson 1983]. In this model, presupposition is not defined by sharp logical criteria. Its resistance to negation and defeasibility is variable and word-dependent.

*My cat is not hungry.*

The existence of my cat is presupposed – whether or not it is hungry, and whether or not the listener knew about it.

In this account, my cat was a latent fact with low prominence, brought into the common ground by shared world knowledge rules such as [PERSON HAS PET] and [CAT ISA PET]. My mentioning the cat makes it an explicit fact with high prominence (whether or not I say it is hungry); the cat survives my negation of its hunger.

*My sunset is cloudy*

This poses problems of interpretation, because people do not own sunsets. Owned sunsets might be metaphorical, in a sub-context of OLD_AGE; or an owned sunset might owe its existence to a sub-context with world knowledge [PERSON MAKES PAINTING] and [PAINTING DEPICTS SUNSET]. Resolution (for both speaker and listener) depends on the relative prominence of different sub-contexts

> *She does not regret doing that.*

The presupposition that she did that often survives negation. For any deed D, both 'she regrets D' and 'she does not regret D' presuppose that she did D. In this model, for both speaker and listener the deed D becomes prominent in the common ground while producing or understanding the whole utterance; and it stays prominent, with or without negation in the whole utterance. This is part of the learned definition of 'regrets', and it is word-specific.

Another example shows that individual words differ in their resistance to negation:

> *John did not realise it was illegal*
> *John did not think it was illegal*

'realise' has greater negation-resistance than 'think'. Similarly, 'warn' has greater negation resistance than 'tell', or even 'allege'. Within 'tell':

> *John did not tell me he was leaving the college*
> *John did not tell me about the roadblock*

Here, the entities 'roadblock' and 'college' have greater negation-resistance than the act of leaving.

Negation resistance is defined in the lexicon. Apparent regularities of negation resistance are scattered across a large lexicon, acquired by profligate learning. As was described in section 9, partial regularities in the lexicon (like language universals) arise by processes of historic language change.

All the inferences described above are equally accessible to speaker and listener, provided their common grounds are initially in step (including that they have the same language constructions and general knowledge).

### Particularised Implicature

Amongst pragmatic phenomena, implicatures are characteristically Gricean – where both speaker and listener assume that the speaker is trying to follow some cooperative maxims.

> *Shall we ask Billy to the film?*
> *He has to finish his essay.*

Here, a speech act ENQUIRY is normally followed by an ANSWER speech act. The first speaker has made an ENQUIRY, and by the cooperative principle, expects to get an ANSWER. (in this model, both interlocutors can gain in status by continuing the conversation; therefore they cooperate to sustain it)

The second utterance qualifies as an ANSWER because of the world knowledge [DOING X] => [CANNOT DO Y] (which has detailed dependences on activities X and Y)

Similar principles apply to

> *Can you tell me the time?*
> *The milkman has just called.*

The second utterance counts as an ANSWER to a time range question because of world knowledge like [MILKMAN CALLS 9_O'CLOCK]. Both speaker and listener need to identify the approximate precision of the time range required for an ANSWER, depending (for instance) on the approximate current time of the question.

Another example is

> *Where's Bill?*
> *There is a yellow VW outside.*

The second utterance counts as an ANSWER because of shared world knowledge like [BILL OWNS VW]. The speaker needs to know that the listener has this knowledge. Facts about VWs are retrieved from the common ground by fast associative retrieval by the meaning VW.

Both speaker and listener need to infer how large (and how useful) is the spatial region required in an ANSWER; they know that 'Bill is in the UK' will not suffice, but 'Bill is on a flight to Australia' might suffice.

> *It is freezing*

One learnt meaning of this whole utterance is [SPEAKER FEELS COLD]. The alternative meaning depends on the prominence of [COLD WATER], so there is a prominence competition between these two meanings – which might be resolved by ostensive shivering, pointing at a container of liquid, or by the preceding utterances.

> *It is raining*

This also might be learned as a whole utterance, but the key fact which needs to be resolved is: where is it raining? Resolving this depends on using world knowledge like [RAIN OUTDOORS], leading speaker and listener to choose the most prominent outdoor location – here, or the nearest outdoor place.

### Generalised Conversational Implicature

Generalised Conversational Implicatures (GCIs) [Levinson 2000] are those which are the least dependent on inference from facts and rules in the common ground, and on the relative prominence of those facts.

In this theory, which depends on profligate learning of large numbers of language rules, the account of many GCIs (such as scalar implicatures) is that they are just a part of the meanings of words; many words are learnt and used to follow GCIs. For instance, in the scale of meanings no/some/all, 'some' usually means 'some, and not all', but that meaning can be overridden to mean 'some and possibly all' – by the use of extra words. Because this is a probabilistic theory of language, the word 'some' can mean 'some but not all' with probability $P > 0.5$; and 'all' with probability $1-P$; and the probability P can be learned from the balance of examples heard. A similar account applies to other scalar implicatures.

To some people, this may seem an unsatisfactory and messy account. GCIs appear to be neat facts, holding across large parts of language. Surely they deserve a neat explanation, and they should not be just a consequence of large numbers of learnt word meanings?

To meet this objection, there is a second layer of explanation. This depends on processes of historic language change, as described in section 9. A language is a large set of constructions for words and multi-word fragments. Language change happens by the evolution and selection of individual constructions, as they are used and as they reproduce by learning. If any construction species is to survive, it must find a viable niche of meaning – in which it is used frequently enough to reproduce (be transmitted) by learning. Scalar implicatures are shaped by the evolution of constructions.

For instance, if the variable S denotes some set of entities, the language rules for the (no/some/all) word species are:

['no', S] ⇔ [EMPTY_SET]         P(B|A) = 1.0

['some', S] ⇔ [SUBSET_OF S]     P(B|A) = 0.7

['some', S] ⇔ [FULL_SET S]      P(B|A) = 0.3

['all', S] ⇔ [FULL_SET S]       P(B|A) = 1.0

By occasionally using a longer expression such as 'some, possibly all', these words can express all the meanings which are ever needed. Just this set of word species has survived in English.

Suppose there was an extra word 'allsome' with constructions

['allsome', S] ⇔ [FULL_SET S]   P(B|A) = 0.7

['allsome', S] ⇔ [SUBSET_OF S]  P(B|A) = 0.3

There would be some occasions where the word 'allsome' can be used instead of the clumsier expression 'some, possibly all', with 'allsome' having its most probable meaning. However, the meaning niche for the word 'allsome' is not viable; people are prepared to use the longer expression on the few occasions when they need to, and 'allsome' will not be used frequently enough to be learned and transmitted. The construction becomes extinct, or it never gets started.

Evolution of constructions tends to lead to linear ordering of meaning intervals (scalar implicatures), rather than more complex structures, such as one interval dividing another interval.

In a similar way, the set of words for colours in a language settles to a size which serves the needs of a speaking community – which divides the space of colours into a useful set of colour regions, which people have a need to talk about. If people rarely need to say 'purple', the construction becomes extinct – or never gets started. Colour word meanings evolve to occupy contiguous areas in colour space.

As another example of a scalar implicature and its violations, consider:

*Joe has three children*
*I have three coins*

The constructions for 'three' are

['three'] ⇔ [EXACTLY 3]         P(B|A) = 0.7

['three'] ⇔ [ >= 3 ]            P(B|A) = 0.3

So when 'three' is used as an ANSWER to an ENQUIRY speech act, as in the first example, the more probable meaning is used. But in the second example, when it is in a part of the act of COMPLYING to a REQUEST 'can you lend me three coins for the parking meter?', and there is a high probability (in world knowledge) that the lender has many coins, and there is no need for the lender when COMPLYING to inform the borrower how many coins he actually has; so the second meaning is preferred, by both speaker and listener.

In this way, processes of historic language change lead to the partial regularities of language, to cross-language universals, and to an account of GCIs.

### Speech Acts

In this model, a great diversity of speech acts (perhaps organized into a hierarchy of types) is not a difficulty. Speech acts existed from the very beginnings of language, as in section 8. If language is joint action in a common ground, then every utterance is an action which alters the common ground. The diversity of languages and cultures leads to a diversity of speech act types, created in group-specific ways. There are culture-specific speech act types such as INITIATION, BLESSING, CURSE, and so on.

In many cases, speech acts follow other speech acts in a defined sequence such as ENQUIRY => ANSWER; or the pairs of act types may be recursively nested [Levinson 1983]. The prominence stack of the common ground gives a direct way to represent nested speech act pairs; the latest nested speech act has higher prominence than an outer (earlier) ENQUIRY, so it will be paired before the ENQUIRY is paired with an ANSWER. Nesting structure of speech acts is supported by prominence.

In most societies there is a speech act type REQUEST for the listener to do some action, and ENQUIRY for the listener to provide some information. In English, these can both be prefaced by 'Can you..', so there is an ambiguity to be resolved by both speaker and listener – which of the two speech acts is intended? As an example:

> *Can you pole vault?*
> *Can you pass the salt?*

There are two learned constructions for 'Can you X?', with different speech act types. How is the type of each speech act to be determined? World knowledge can be used to determine whether an action is feasible, and whether it benefits the speaker - implying a REQUEST for action, rather than an ENQUIRY for information.

In the first example, there is world knowledge [POLEVAULT USES LONG_POLE] and [POLEVAULT NEEDS SKILL]. Although there is also weak knowledge that sometimes [POLEVAULT BENEFITS SPECTATOR] (because SPECTATORS enjoy watching ACTs), in most contexts there is no prominent entity LONG_POLE in the common ground, so the action is not feasible, and the possible benefit to the speaker is not enough to make it a REQUEST; therefore it is understood as an ENQUIRY.

In the second example, there is usually a prominent entity SALT_CELLAR in the common ground, in an accessible location; and for many types of food, there is world knowledge [FOOD NEEDS SALT], which together with the prominent fact [SPEAKER EATS FOOD], establishes both the feasibility and the benefit to the speaker, making the speech act a REQUEST for action. If the addressee is a small child, the feasibility of the action is not so clear-cut, and the speech act might be ENQUIRY for information, about the child's capabilities. All these inferences are available for both the speaker and the listener.

For any REQUEST, the speaker and listener need to agree the (approximate) range of times over which a REQUEST is to be fulfilled. For 'pass the salt' this typically depends on world knowledge in the MEAL cluster. Other REQUESTs may allow a much longer time for fufilment.

In another example

> *How do you do?*

there is a simple learned language construction

['howdjado'] ⇔ [GREETING]

used by the speaker (right-to-left) and the listener (left-to-right).

In most cases, if the listener s to correctly infer the speech act type, he must correctly infer what the speaker intends; he must infer something about the speaker's state of mind.

Speaker states of mind can be described by nested Theory of Mind feature structures such as

[S INTENDS [X]]

Or

[S BELIEVES [X]]

Where X can be any meaning. These ToM feature structures can be the semantic pole of pragmatic constructions; such pragmatic constructions can be learned from a few examples, like any other part of a language. Once learned, a ToM construction enables a listener to infer something about the speaker's state of mind, by a rapid, pre-conscious unification operation. This capability gives humans a fast Theory of Mind ability.

For instance, to distinguish between REQUEST and ENQUIRY speech acts, people learn ToM constructions enabling them to determine whether the speaker wants some new information (ENQUIRY) or some other act performed (REQUEST).

### Irony

An ironical tone or facial expression means both 'some people P believe Y' and 'I do not believe Y'. Both are represented by a nested Theory of Mind feature structure [X <NOT> BELIEVES [Y]], where X = some people P (without the optional NOT) and X = SPEAKER (with the NOT), and the nested feature structure Y represents the proposition which is believed (or not believed). The set of people P who believe Y needs to be identified in the common ground, by both speaker and listener.

> *Yeah, Sure!*

Here, the inferred proposition is one very recently expressed. The inferred believer P is either the previous speaker, or somebody he has made prominent in the common ground (e.g. by saying 'Fred thinks..). The non-believer is , of course, the speaker.

> *She is a real paragon of virtue!*

Here 'paragon of virtue' started as a metaphor, but is now used and learnt with a frozen meaning [VERY_VIRTUOUS PERSON]. 'She' is the most prominent female person in the common ground. Emphasis on the word 'she' indicates irony, and it still needs to be inferred what group of people P believes the inner proposition about her virtue. This is some group or person recently made prominent.

### Hyperbole

The account of hyperbole may be straightforward – that speakers often exaggerate (for instance, to make a point more clearly, or to flatter) and listeners learn not to believe them – to dilute the meaning when appropriate. This applies to

> *You're a genius!*

In simple cases, words simply change their meaning. For instance, 'awesome' (when applied to most peoples' actions) has recently come to mean QUITE GOOD, and 'genius' has long meant QUITE CLEVER rather than EXTREMELY_CLEVER – except in rare contexts of exceptional scientific or artistic achievement.

Again, a pragmatic device is mainly a matter of learning the meanings of many words – including different meanings for the same word in different contexts.

### Metaphor

Novel metaphors may require novel inference, possibly including mental simulation. After a few examples, the metaphorical meaning is learnt and applied by a single unification. As an example of a novel metaphor:

    A. *Will John look after our interests in the faculty meeting?*
    B. *John is a soldier.*

Hear, speaker B chooses a metaphor, and needs to ensure it will not be misunderstood by A.

To choose a novel metaphor, B has world knowledge about a range of professional roles – TINKER, TAILOR, SOLDIER, SAILOR, and so on. For each role, she has several pieces of world knowledge, such as [SOLDIER HAS RANK], [SOLDIER WEARS MEDALS], [SOLDIER TRAINED_FOR ARMED_COMBAT], [SOLDIER OBEYS ORDERS], [SOLDIER LOYAL_TO ARMY]; similarly for the other roles.

B is seeking a metaphor which expresses that John is loyal to our group (which is a research team in the faculty), and will try to influence faculty meeting decisions to benefit our group. B can associatively retrieve all roles which have the attribute LOYAL_TO some group. This picks out the role SOLDIER, but not the others; TINKERS are not known for their loyalty. So the choice of role for the metaphor can be done by fast associative search, not requiring slow serial search.

To ensure that the SOLDIER metaphor will be understood correctly, as implying LOYAL_TO, and not some other attribute of soldiers like ARMED_COMBAT, the speaker needs to check each well-known attribute of SOLDIER against the most prominent facts in the common ground – in this case, facts about FACULTY_MEETING. Her world knowledge includes the fact that faculty meetings take decisions by DISCUSSION, not by ARMED_COMBAT; so that this misunderstanding of the soldier metaphor cannot arise. In the case of RANK, there may be a greater risk of ambiguity; decisions in faculty meetings are swayed by the RANK of the proponent. The test for ambiguities is a two-way match between the attributes of SOLDIER and the attributes of FACULTY_MEETING. The brain may have some fast way to do this; but generally, to check out a new metaphor, flexible general inference may be required; or occasionally, there may be a need for repair.

The listener does not need to consider any alternate metaphors such as TAILOR; but, like the speaker, needs to do a two-way match between SOLDIER and FACULTY_MEETING. He arrives at the intended meaning LOYAL_TO; or (if he has a different experience of faculty meetings) may choose an unintended sense RANK.

Suppose the second speaker had made a different reply:

    B. *John is a Trojan horse.*

This is no longer a novel metaphor. At one time, people with some classical education would have understood it by recalling the story of the Sack of Troy, and role of the soldiers in the wooden horse. The metaphor has now come to mean:

['Trojan horse'] ⇔ [LOOKS_LIKE BENEFIT & ISA THREAT]

This is the only sense in which the phrase 'Trojan horse' is ever used, so this meaning is simply learned as a construction, and applied by simple unification (again, right-to-left for the speaker, and left-to-right for the listener). No other inference is required by either interlocutor.

Were it not for a cultural accident, the same metaphor might have acquired the meaning

['Trojan Horse'] ⇔ [CANNOT JUMP & CANNOT RUN]

In other words, a Trojan horse cannot do the things it was intended to do, and is useless. If fresh inference was required on every occasion the metaphor was used, listeners might frequently make this interpretation – which they do not. The metaphorical meaning USELESS has been learnt as a single construction.

Similarly the metaphor

['chocolate teapot'] ⇔ [NON_WORKING THING]

requires physical imagination when it is first encountered; but can then be learned as a construction.

In summary, novel metaphors may require flexible inference or repair; but once learned, a metaphor is simply a normal language construction.

Spatial metaphors have a special place in language. This is because concrete physical meanings can be consciously experienced by imagining them (imagine a dog). Abstract meanings cannot be experienced in that way, unless they use some spatial metaphor. Spatial metaphor lets us consciously imagine some part of an abstract meaning,ardin (for instance) consciously simulate some of its consequences. For this reason, spatial metaphor is very widespread in language [Lakoff & Johnson 1980].

Here are two examples of spatial metaphor:

    *Cheer up!*

At some time, this may have appealed to a meaning like: 'be of good cheer; look up to the heavens, not downwards towards hell'; or by using the fact that when someone is unhappy, their posture is downward-looking. However, the phrase has come to have only one meaning, as a speech act of type REQUEST (or COMMAND):

['cheer up'] ⇔ [REQUEST: BE CHEERFUL]

So it is simply learnt and used with this meaning; its metaphorical origin has been almost lost.

*Get down to it*

Its original meaning may have been that to do some extended task, you need to sit down, or look down at some piece of work. However, it is always used with meaning

['get down to' TASK:T] ⇔ [INITIATE T]

where the task T ('it') needs to be identified in the common ground – perhaps from some recent utterance, or by pointing at a piece of work to be done. Often 'it' is sufficient to denote the task T.

Hence there is no need to parse the words 'get down to' or to construct any literal 'semantic' meaning from them. The utterance 'get down to it' is learnt as a whole piece (as one construction), and simply applied by unification.

In summary, just as language syntax and semantics are described in construction grammars by fast, pre-conscious unification of constructions, similarly pragmatic phenomena can be understood by fast unification of pragmatic constructions, including facts in the common ground. There is no boundary between pragmatics and semantics [Ariel 2017]. Many pragmatic constructions are Theory of Mind constructions, impelling us to make rapid inferences about other peoples' states of mind – taking the 'intentional stance' of [Dennett 1989].

## 13. The Major Puzzles of Pragmatics

This section starts from a survey of the major outstanding issues of pragmatics, in Stephen Levinson's [2023] book 'The Dark Matter of Pragmatics'. It asks: does the current theory offer possible solutions to those puzzles?

Under five major headings, Levinson discusses progress in the study of pragmatics - questions he regards as relatively well understood – and the outstanding puzzles, which he calls the 'dark matter' of pragmatics. Finally, he discusses an overarching puzzle which he calls the 'dark energy' of pragmatics.

This survey, by a founder and recognised leader of the field of pragmatics, is an appropriate framework to address its main issues – to survey the wood, without getting lost in the undergrowth.

Levinson identifies a 'pragmatic gap' between the information rate we achieve by speaking (about 40 bits per second), and the higher information capacity we can receive as listeners/observers. It is this gap which contains the dark matter of pragmatics, and he frames the five puzzles in terms of possible engineering solutions, to bring us closer to the maximum receiving bandwidth.

The first question to ask is: why does communication bandwidth need to be maximized? The fitness benefits of great speed, in a natural habitat, are marginal compared to the metabolic costs of a large brain. This theory answers the question by sexual selection for language: because speed is impressive; the intention game requires us to be impressive, to show off our intention reading ability in order, ultimately, to pass on our genes. Sexual selection has been a strong selection pressure to drive our language ability and high bandwidth.

I shall briefly describe each of Levinson's five dark matter puzzles, describing in each case how this theory of language can solve the puzzle.

### Puzzle 1: Multi-modal packaging:

Levinson's first bandwidth-enhancing solution is to convey information by multiple channels, such as facial expression, gaze, or gesture. [Levinson & Holler 2014] The resulting puzzles are: how does a speaker encode information in these multiple channels, and how does a listener decode it?

In this theory, language originated in multi-modal interactions; while the voice channel later came to dominate, other channels have not disappeared, because they assist fast communication. The channels are integrated by **fast unification** of constructions, which embody learned constraints between the timings of different channels – for instance, a gesture must precede a word to convey a particular meaning. A speaker produces language by using constructions whose phonological pole includes not only sound sequences, but also other channels like gesture and facial expression and intonation (which were all used in earlier stages of the intention game, and never went away). Feature structures in constructions can include any of Levinson's [2023] 150 signaling devices. These have fuzzy learnable constraints on their relative timings, like: 'the delay between the word and the gesture is between 0 and 100 ms'. These fuzzy, probabilistic constraints fit easily into a Bayesian model of processing.

Speaker and listener have the same constructions (by the learning guarantee); and by unifying the feature structures in different orders, the listener can reconstitute the speaker's intended meaning (by the communication guarantee), binding together all the multi-modal cues. Unification is a fast, pre-conscious, parallel operation, which can happen at rates of 5-10 unifications per second. Fast unification is the solution to the multi-modal binding problem.

### Puzzle 2: Speech Acts

Bandwidth can be increased if each utterance does double duty, both as a communication and a deed – a speech act. The puzzle is 'how actions are mapped onto words'; how do participants actually ascribe actions to utterances? There are subsidiary puzzles – is there a finite list of core action types? Is the list of types cross-cultural?

In this theory, speech acts pose no great difficulty, but equally there is no single neat solution. The earlier stages of the intention game were all about social actions; within one hominin group, we would expect to see an increasing number of learnable specific actions, with broad classifications of their types (possibly as feature structures higher up in an inheritance hierarchy) emerging over historic time in a group-specific, culture-specific manner. As the intention game grew more complex and demanding, all the communication tools (gesture, speech, ..) were in effect actions. While they grew in semantic complexity as our mental powers grew, they retained their action component; and they retained their dependence on facts in the common ground (context).

Through all these stages of the game, fast unification remained the standard tool for interaction – because speed remained of the essence, and the communication guarantee remained necessary. That is part of the reason why the attribution of speech act types is fast and apparently effortless.

When A used some pragmatic construction to encode information and an action, and B used the same construction to decode it, this automatically activated constructions above that construction in the hierarchy, which hold less information – essentially defining broader types of action. To that extent, they are speech act types. These broad types would have some universality, which (like other language universals) emerged over historic time, time and time again in different groups, by processes described in section 9. This happened from the early stages of the intention game, and continues to happen today. The range of speech act types tell us less than we might think about the human brain, and more about historic language change.

Another puzzle here is how participants are able to attribute speech act types early in the comprehension process, to start planning their fast turn-taking. Early in comprehension, a listener can navigate far enough down the feature structure hierarchy to start planning his own response. To some extent, this may be a product of the way unification-based understanding works, passing some signal down the hierarchy; but it is also a result of how constructions have evolved over historic time. There is a selection pressure on pragmatic constructions, to make them partly understandable early, avoiding hard ambiguities and defining a broad speech act type for early response planning. Constructions with hard ambiguities are selected against, and go extinct. Constructions evolve very fast under this selection pressure, within a few human generations – as described in section 9.

The challenge of finding a speech act type is a problem in mind-reading – what does the speaker intend to happen next? Since the very beginnings of the intention game, the relevant pragmatic constructions have been Theory of Mind Constructions – cementing the union between language and mind-reading, and giving our fast pre-conscious mind-reading ability.

That is a brief sketch of the answer this theory gives to the problem of action ascription – which Levinson calls '*the beating heart of language use – the most vital part of pragmatics*'. It is mind-reading done by fast unification, depending on the communication guarantee and the historic evolution of constructions.

**Puzzle 3: Message Form and Compression**

Under this heading, Levinson discusses issues such as conversational implicature and presupposition – how the form of a message conveys extra information beyond 'what is said'.

Much of this problem has been created by trying to draw a hard boundary around 'what is said' – trying to work out some logically-based semantics, before tossing it over the wall to pragmatics; and turning a blind eye to 'context' before you toss it. Much effort has been spent debating where the boundary should be drawn. This theory has no boundary between pragmatics and semantics. If there is no boundary – if pragmatics and semantics can overlap freely in the order of unification, much of the problem disappears.

I illustrate this by one example of implicature [Wilson 2017]:

- *Shall we ask Billy to the film tonight?*
- *He's got a paper to finish*

In relevance-based approaches to pragmatics, the listener has to do expensive, open-ended inference, to find out the relevance of Billy's paper to her enquiry about the film. She has to find out what is the implicature, which makes the second statement count as a reply? (the speaker has to do even more complex inference, which has not been discussed in relevance theory)

In this theory, the first time a listener encounters this kind of reply, there is an expensive slow inference to do; and if that inference fails, she may need to ask for repair. After encountering a few examples of this kind of 'reply', she encodes them in a feature structure – which may initially be quite specific:

X has to go to bed => do not ask X to play

Later, this learned feature structure may be further generalized; if X has to go to bed, do not ask X to do anything else – play, or come out for a walk, or listen to a story…. The next stage of further generalization is

X has to do Y => do not ask X to do Z

with encoded restrictions on Y and Z; Y is time-consuming, Y needs to be done soon, Z needs to be done soon, and so on.

Equipped with this learned feature structure, in the Billy film example the listener no longer needs to do expensive open-ended inference to establish the relevance of the reply. A bare semantic representation of 'what is said' is not created, because many words in the language have pragmatic implications; they may be unified at any time in the unification order. It is a simple fast pre-conscious unification; she just 'knows' her question is answered, with no conscious effort. She can understand that it is a reply, having heard only the first three words "He's got..". Not only that – she can use the same feature structure in reverse, to answer questions herself. The communication guarantee ensures that the exchange will work.

A similar analysis is required for presuppositions – here it is more closely linked to other language learning. When the child first encounters the word 'regrets', she needs to painstakingly analyse the surrounding words and the common ground, to infer that:

a) 'regrets' implies a mental attitude to some act – of wishing it had not happened
b) Usually, the act has happened
c) Sometimes (b) is 'cancellable'; something else in the common ground or the utterance implies that the act has not happened.

This kind of slow inference from the common ground is part of the process of learning the feature structure (or feature structures) for any word or construction in the language – whether or not it carries a strong presupposition as 'regrets' does. Without a semantics/pragmatics boundary, presuppositions are simply learned as parts of word meanings; different strengths of presupposition are all stored in the lexicon.

A similar analysis applies to the contrast between 'he stopped the car' and 'he caused the car to stop'. You learn from examples that 'stop the car' refers to the usual 'hit the brakes' act; whereas 'cause' is an open-ended way of saying 'event A leads to event B', and is used in cases where the link from A to B is not the obvious default one.

Similarly, scalar implicatures [Horn 1984] are part of the meanings of words, to be learnt. Feature structures have the ability to store fuzzy slots, with ranges of possible or likely values; the fuzzy ranges are learnt by generalizing examples. The meaning of 'some' has a fuzzy slot implying 'usually not all', but in Bayesian inference, this can be overridden by other information. Many adjective meanings, like 'big' are intrinsically fuzzy and relative. This may seem to require a powerful model of learning, but even small mammals do it all the time [Anderson 1990]. The Bayesian nature of language constructions replaces many problems, which previously appeared intractable in a discrete logical framework, by problems of Bayesian probabilistic inference – to find the most likely meaning of an utterance.

So the answers to many puzzles of implicature and presupposition are broadly ones of powerful learning, and Bayesian probabilistic processing; but this needs to be worked though for examples, to get a feel for how it works.

### Puzzle 4: Non-Literal uses of Language

Here, the puzzle is that we have '*no real algorithms*' for how non-literal uses of language such as metaphor are processed: '*given the cascade of inferences potentially triggered…the speaker's target inferences need to be narrowed down..*' Levinson may be seeing this problem primarily from the listeners' point of view; but the inference problem for the speaker is, if anything, more intractable: '*If I say that, will she have sufficient clues to work out what I mean?*'.

For this theory, the solution to the problem lies again in painstaking slow inference for the first few examples of any pragmatic device, learning a pragmatic construction from those examples, followed by fast, pre-conscious unification for all later cases. In Levinson's terms, that is a 'real algorithm'. It works, for both speaker and listener, through the communication guarantee. We know how they share the same feature structures, because of the learning guarantee.

As was described in an earlier section, it is likely that many tropes of communication, including metaphor and irony, date back to the earlier stages of the intention game. In the early stages of the game, before we had spoken language, we needed to be skilled at these tricks, to be impressive intention readers and to pass on our genes. With the advent of spoken language, these tricks did not disappear from our communication toolkit.

There is a particular role for spatial metaphors [Lakoff & Johnson 1980]. From the earliest stages of the intention game, conscious reasoning has been required to understand and learn some new communication device as a feature structure. Because our consciousness is largely spatial – a 3-D conscious model of the world - conscious reasoning means spatial imagination and simulation; so metaphors, which allow non-spatial meanings to be mapped onto spatial meanings, are a vital tool to extend the range of things we can communicate about, to show off our intention reading. Anyone incapable of learning spatial metaphors would be regarded as a loser in the game, and have a hard time finding a partner.

### Puzzle 5: Leveraging the Context

For Levinson, the '*black hole of context*' is that there seems to be '*no end to the factors that might be relevant to understanding a talk exchange*' and '*we must swim in a vast sea of attributed beliefs*'. He sees the great difficulties for speakers and listeners, as being difficulties for a theory of pragmatics.

If we are looking for clean, elegant solutions, with crisp boundaries between semantics and pragmatics, there may be difficulties for the theorist. There would be a lot of work to do, to construct any elegant theoretical model. But is that what we need to do? Do we need to find neat semantic rectangles in the river of language?

The 'black hole of context' is precisely the common ground – which was the foundation for language, long before there were spoken words or syntax. Language has always required us to understand the common ground, including other people's states of mind. That has always been the driver of the intention-reading competition.

As Levinson says, there is "*no end to the factors* [in the common ground] *that might be relevant*"; the common ground potentially contains an infinite number of facts. However, as described in section 8, most of these are implicit facts, generated by rules; and all facts in the common ground are ordered by their prominence, or expected relevance. So the search for the correct interpretation is an ordered finite search. We have become very good at it, through strong sexual selection.

In the early stages of the interaction game, most of what the players needed to deal with was the context, or facts in the common ground; only in the later stages of the game did the balance shift towards new facts introduced by the players. In later stages of the game, players evolved to have an increasingly prodigious ability to learn and unify feature structures describing facts in the common ground, and the facts they added to it.

This was not designed to be an easy game to play. Over evolutionary time, the intention game needed to become harder and harder – to be an ever more stringent test of the players' intelligence, to enable them to size each other up as their ability increased. It should not be a surprise that there are no easy criteria to define what type of facts in the common ground may be used. The competition was not designed to be easy.

In this theory, we should not be looking for elegant general solutions to the problem of context. When a hominin group devised some new way to exploit the common ground in their intention game, nothing was off limits; any type of fact could be used. The first time someone tried out some new move in the intention game, and someone else observed it, it required slow general-purpose inference to work out what was meant. Then, after even a few examples, the move could be encoded as a learnable construction. (if it could not be encoded, it would not become part of the game; it did not even start to evolve a construction species) The players effectively agreed that this new move was now part of the game, and used it by fast unification. Anybody who could not use the new move was marked down as a loser; the purpose of the game is to identify losers.

Over time, each new move would itself be subject to natural selection, as in section 9; a hominin group would have a limited set of moves they regularly used, and constructions would be selected to be part of that set. If a move in the game was not regularly used, it would die out. It could only be regularly used if it could be encoded as a feature structure, which could be used by unification on a reasonably large number of occasions, obeying the communication guarantee and the learning guarantee. This was naked competition between feature structures; but having a simple meaning, using only some restricted set of facts, was not the only competitive asset. A feature structure species needs to tread a tightrope, between being difficult enough to be a good test of intention reading, and being usable on a reasonably large number of occasions – enough occasions to be propagated by learning.

In sum, as theorists we should not be looking for simplicity or elegance or restrictions in exploiting the common ground. As people became smarter, they needed these constraints less and less; so the evolution of constructions has led to ever less constrained pragmatic constructions, to be a more stringent test of intention reading.

### Puzzle 6: The Dark Energy of Pragmatics

Levinson borrows the term 'dark energy' from cosmology to mean '*whatever it is that propels our pragmatic system at the incredible speed it seems to operate*'.

By now, the answer from this theory should be familiar. The dark matter, which allows any pragmatic inference to be made rapidly and pre-consciously, is unification of feature structures. From fast parsing of elementary syntax and semantics, we know that the human brain can selectively retrieve and unify up to 10 feature structures per second, whether in production or comprehension – out of a stock of many thousands of learned constructions. All that is required is to extend this capability to pragmatic inferences, which extends the communication guarantee and the learning guarantee.

That, in a nutshell, is the answer to a central mystery of pragmatics – our ability to take our conversational turn within 200 ms of the other speaker finishing their turn [Levinson & Torreira 2015]. We do it by finding and pre-consciously unifying just the right feature structures, to start planning our own turn well before the other person finishes their turn.

In defining the dark energy of pragmatics as the '*incredible speed*' of our pragmatic inferences, Levinson does not say whether he is referring to the inferences required for pragmatic comprehension, or those required for pragmatic production, or both. It appears that in common with most published work on pragmatics, he is mainly thinking about the inferences required by a listener. Little attention has been paid to the inferences required of a speaker.

The listener has to deduce: 'In this context C, from the heard words W, what meaning M does the speaker intend me to infer?'. The speaker has to deduce: 'In this context C, I intend the listener to infer only the meaning M. What words W should I use?'.

At first sight, these are two different problems of inference. As a problem of trial-and-error search, the speaker's problem appears to be the harder of the two:

- The listener has to search over a number of possible meanings M and M', finding one that is consistent with the words W, and has sufficient relevance to the context C.
- The speaker has to internally try out some set of words W, then reproduce the listener's whole inference process, to test whether it results in the meaning M. If it does not, the speaker has to try out a different set of words W', until he finds the right words to say, so the listener will infer the right meaning M.

It is clear that both speaker and listener need some clever guidance to restrict the possible search space – the speaker, more than the listener, because the speaker's search may be more time-consuming. The speaker needs to internally reproduce the listener's whole inference process, and possibly do it several times.

In this theory, the clever guidance is provided by fast associative retrieval of the appropriate feature structures (by their meanings or by their sounds), followed by fast unification – which results in the communication guarantee, that the listener can recover the speaker's meaning M. This drastically reduces the search space for both speaker and listener. It implies that the speaker never needs to literally reproduce the listener's inference process. The theory provides a working computational solution to the problem of fast inference, for both speaker and listener.

I know of no other proposed computational solution to this double inference problem; that is why it can be called supreme 'dark energy' problem of pragmatics. Solving this double problem is perhaps the main claim of the theory.

### The Historic Neglect of the Speaker's Inference Problem

The study of pragmatics has focused almost entirely on the listener's inference problem, of recovering the speaker's intended meaning – while almost neglecting the speaker's inference problem, of 'how to say it?'. The reason may lie in history of pragmatics, which started in 1957 with Grice's (1989) maxims.

An unstated reason for this systematic neglect may be something like this: 'Empirically, speakers follow Grice's maxims. That is not a problem. There is a difficult inference problem for listeners, in recovering the speaker's intended meaning. A listener can assume that Grice's maxims have been applied, to constrain their inferences?'.

The incompleteness of this approach can be seen by imagining a Grice's maxim for footballers:

*If your team is losing the game, score a goal.*

Surely the rules of football define what a goal is, perfectly clearly. Where[2] is the problem?

One of Grice's maxims starts with the words 'Try to'. If all of Grice's maxims were prefaced by the words 'Try to', that would scarcely alter their force and meaning. This shows that all of Grice's maxims define goals for the speaker – but they are not prescriptions for how to attain the goals (to attain all the goals in as short a time as possible; and not to violate any of the goals in any utterance, unless 'flouting' a maxim for some purpose). Doing so requires inference which, as argued above, is harder than a listener's inferences.

In neo-Gricean pragmatics, Horn's (1984) Q- and R- principles, and the speaker's maxims in Levinson's (2000) Q-, I- and M- principles, can all be prefaced by the words 'Try to', making it clear that they all define goals for the speaker; they are not 'how to' prescriptions to achieve those goals.

In relevance-based pragmatics [Sperber & Wilson 1986,1995, 2012], I have seen little mention of the speaker's inference problem. [Wilson 2017] says '*the challenge for relevance theorists …. is precisely to explain how the closed formal system of language provides effective pieces of evidence which, combined with contextual information,* **enable successful comprehension to take place**' – thus excluding the speaker's inference problem of production.

Wilson goes on to cite the relevance-theoretic equivalent of Grice's maxims – the communicative principle of relevance:

*Every utterance communicates a presumption of its own optimal relevance*

The definition of 'optimal relevance' has two legs – loosely, that the communication is 'relevant enough' and 'the most relevant possible'. Both of these are used to define '*what the addressee is entitled to assume*' for his inference; he can assume that the speaker is trying to achieve optimal relevance, and that the speaker succeeds. This side-steps the speaker's problem.

As described above, the speaker's problem seems to include solving the listener's problem – possibly several times over.

---

[2] The simple injunction to 'score a goal' hides a problem of nested inference, similar to the speaker's inference problem. You need to think of several possible attacks. For each attack, you must mentally simulate how the defence will respond. Only when you find an attack that cannot be defended against, can you do it.

It must be the harder of the two. Levinson's dark energy puzzle is twice the puzzle it appears to be; but this theory offers a solution, in fast unification.

## 14. Consciousness and Language

Much of this paper has been devoted to the fast, pre-conscious operations of language – particularly the many fast retrievals and unifications of learned constructions, which we need to do to parse and build the meaning of any utterance we hear, before we can be consciously aware of the meaning. These unifications all happen outside conscious awareness.

However, one of the most important facts about language is that we are consciously aware of language meanings, for instance as mental images. Any complete theory of language needs to give some account of this fact.

A major difficulty in this respect is that our current theories of consciousness are so many and so unconvincing. It is not useful or possible to try to relate all these theories to language. I shall start with some evidence about consciousness, and then relate language to one theory of consciousness, noting how the same model of conscious language might apply more broadly.

Evidence about consciousness is in the front of our minds at every moment of the day. A large part of our everyday consciousness consists of spatial consciousness – a three-dimensional conscious model of the space around us, and all the things in it, with ourselves at the centre. A first task for any theory of consciousness is to explain how spatial consciousness arises.

Most theories of consciousness fail to do this, because they assume that consciousness arises only from neuron firing, and temporal neural firing patterns can only hold spatial information in an encoded form. The theories give no account of how neural spatial information in the brain is decoded to give conscious experience.

A theory of consciousness should not only account for consciousness of local spatial reality – it should also account for consciousness of imagined space. This is important for consciousness of language, because language meanings are consciously imagined. In general, imagination is somehow weakly superimposed on consciousness of current spatial reality.

All animals need to maintain a good model of local 3-D space in their brains, in order to plan and control their physical movements. It can be argued that spatial cognition is the most important function in an animal's brain, so it is no surprise that consciousness is linked to it. The model of space which gives rise to consciousness might be stored in the brain in some special analogue way, without encoding – for instance, as a wave excitation in the thalamus [Worden 2024]. Or it may be stored in other ways; we do not know yet.

However the conscious model of space is represented in the brain, it would be useful for any animal to be able to use imagination in the model, as well as actual current space, so it can use the model to imagine (or simulate) possible actions, to find out their possible consequences without the cost and risk of actually doing them. Hence we expect animals to be capable of conscious spatial imagination.

To consciously imagine a possible movement, an animal brain needs to take an unconscious mental representation of a physical movement (for instance, represented as a feature structure graph) and feed it into the conscious model of space, however that is represented.

Hence, we expect that most animals are able to convert feature structures into conscious imagination. Before language, humans used this ability to imagine and plan intended physical actions. This ability could then have been co-opted for human language. In the earliest evolutionary stages of language – the early intention game of section 8 – people used mime and gesture to convey information about their intended actions. To understand a mime, it was necessary to convert a feature structure for the mime to a feature structure for the mimed action, in order to imagine the action. This would have been a fairly simple conversion (because the hand movements can be similar to a mimed action), allowing people to imagine the first language meanings. Then, as vocabulary increased and constructions became more abstract in their phonetic poles (e.g. containing arbitrary sounds as well as gestures), it became necessary to learn the mappings from an abstract phonetic pole to a physical semantic pole – but the task of consciously imagining the semantic pole would become no more difficult.

With the arrival of syntax, the semantic pole of some constructions needs to become a general recipe for constructing a spatial model of a complex situation – for instance, a recipe to build a conscious spatial model of a dog drinking, by combining the semantic poles of the constructions for 'dog' and 'drink'. This problem of combining recipes to build spatial models of concrete things and concrete actions makes extra demands on the operation of unification, but we can begin to see how it might be done. Given the need to imagine meanings in order to carry on conversations, and the intense selection pressure of sexual selection, evolution would find a way.

In this way, we can see how language has evolved to allow us to imagine the meanings of concrete, spatial utterances. More abstract meanings present a further challenge, if the only way to consciously understand them is to build a conscious spatial model. I propose that the main way we consciously imagine abstract meanings is to build spatial models of them. This is the reason why so many abstract words use spatial metaphor in their meanings – to enable us at least to partially understand the meaning, by building a spatial model.

There is another important way in which we become consciously aware of language meanings, both concrete and abstract. I have described how many of the constructions of language are Theory of Mind constructions, allowing us to infer something about another person's state of mind. In conversation, we often need to infer what the other person is thinking about us. In particular, we often infer whether they regard our self as being of higher, or lower, social status. These pre-conscious inferences of status constitute our self-esteem. Self-esteem is an important driver of our emotions – which we consciously perceive as feelings located in our bodies, situated in our conscious model of local space. Hence, we often consciously perceive an emotional impact of something which is said to us (or even of something we say).

These conscious experiences of our bodies result from overt conversations, when we talk to other people and hear them talking. The same processes go on when we mentally rehearse conversations, as verbal thoughts. Mental rehearsal of conversations has evolved to help us make our conversations more impressive, in the intense competition for social status and for mates. Sexual selection has made us avid thinkers; and any thought can feed into our self-esteem and consciously felt emotions.

Therefore we experience verbal thoughts in three ways: as consciously imagined sounds of the words, as conscious spatial images of the meanings, and as emotional feelings in our bodies. This is how conscious language impinges on our minds, for most of our waking lives.

## 15. Comparing the Unified Theory with Evidence

I have now described all the key elements of the unified Bayesian theory of language - and along the way, made comparisons with empirical data about language. The theory claims to describe all the important facts about language. It is now time to circle back, and recapitulate all the comparisons of the theory with data – to check that the theory really can account for the major facts of language. Recall the four main groupings of evidence about language, introduced in section 1:

1. **Descriptive**: Does the theory describe the known features of language?
2. **Computational**: Does it give an adequate account of how language is computed in the brain?
3. **Evolutionary**: Does it give an account of how language evolved in the human species?
4. **Unity**: Can these all be explained with sufficient economy of hypothesis, in a unified and coherent theory of language, rather than a collection of unrelated pieces?

Human language is a huge and multi-faceted phenomenon. In this summary section I cannot do justice to all the facets of language; I can only point out how the theory accounts for some major aspects. Being a unified theory, its account of any language phenomenon may cut across three dimensions of the theory:

   a. its evolutionary dimension, describing the evolution of language in the human brain;
   b. Its cognitive and computational dimension, saying how language works in the brain
   c. its social and historic dimension, saying how languages change over historic timescales .

This table compares the theory with the facts of language, relating them to these three components.

| Feature of Language (Descriptive) | Evolutionary Component (sections 4, 6, 7, 8) | Cognitive/Computational Component (sections 3, 5, 10-14) | Social/Historical Component (section 9) |
| --- | --- | --- | --- |
| Language can convey a vast range of possible meanings, very fast. | Language evolved as a competitive display of intelligence, in competition to get a mate. This has been the strongest recent selection pressure on the human brain. The competitive test of intelligence includes competition for speed, and for expressive power. So language is prodigiously powerful.<br><br>(In natural settings, the fitness benefits of prodigious speed and expressivity are much less than the reproductive competitive benefits. Language is over-engineered) | Construction grammars can express an unlimited range of complex meanings in learned constructions (feature structures).<br><br>Language production and understanding are done by unification of constructions – a fast, pre-conscious operation.<br><br>Fast associative retrieval of constructions enables fast production and understanding of any language, which typically has many thousands of constructions. | In different human social groups, the constructions of language have evolved independently to serve the competitive communication needs of each group. |
| The world's languages have complex and diverse syntax | Using complex syntax is a test of intelligence, and helps to express diverse meanings rapidly; speed is another competitive test. | The constructions of construction grammars can describe the syntax of any language.<br><br>Repeated unification converts a sequence of sounds (and facts in the common ground) into a meaning tree – or in the reverse direction.<br><br>Unification is fast and pre-conscious. | The constructions in each language have diverse syntax, shaped independently by rapid evolution of constructions in that social group.<br><br>The reproduction of constructions by use and learning is preecise enough to support the evolution of complex constructions. Like DNA replication, it has a low error rate. |
| Any language has a communication guarantee – that if two speakers have the same set of constructions, a listener can reconstruct the same meaning that a speaker used to produce an utterance. | Without a learning guarantee, conversations would not be viable – requiring too much repair to serve their selective purpose.<br><br>The communication guarantee was a feature of the earliest multi-modal pragmatic exchanges of the intention game, and has persisted as language became more complex and expressive. | The communication guarantee can be proved from the mathematical properties of repeated unification, as used for language production and understanding.<br><br>The guarantee is also demonstrated in online computational models.<br><br>Because of the communication guarantee, throughout a conversation, participants can keep their versions of the common ground in step, minimizing the need for conversational repair.<br><br>The mathematical basis of the communication guarantee is described in the appendix. | |
| Language is robust, e.g. against missing, mis-heard or unknown words | Language evolved from Bayesian animal cognition, which is a robust optimal way to choose actions in the face of uncertainty.<br><br>Language needs to be robust, to support the cooperative extended conversations which we use to show off our intelligence, even in non-ideal and noisy circumstances. | Bayesian unification in construction grammars finds the Bayesian maximum likelihood parse of an utterance, and the most likely meaning. | If any construction in a language leads to ambiguities too frequently, the construction will not be used, so it cannot reproduce by learning, so it becomes extinct. |
| Ability to infer other people's mental states is a central language skill – particularly for learning, and conversation. | The ability to infer other peoples' intentions from their actions was the starting point of sexual selection for intelligence, in the earliest precursors of language (the intention game).<br><br>From then on, all language has been based on a common ground of mutually known facts. We are selected for ability to understand the common ground, which includes other people's mental states. | To extend construction grammars to describe pragmatics, unification of constructions is unification with facts in the common ground, and is used to infer speakers' intentions as speech act types.<br><br>Language is learnt by inferring, from partially understood utterances, what the speaker is referring to in the common ground. | Theory of Mind constructions – inferring other peoples' mental states from diverse evidence - are needed for pragmatic skills, so they survive the evolutionary selection of constructions. |
| Children learn their native language early, from a few learning examples for each word or multi-word construction. | Language learning evolved from a Bayesian learning ability in other animals, who can rapidly learn very large numbers of regularities.<br><br>People need to start practicing language as early as possible, to become proficient conversationalists before adulthood. So they learn language in early childhood, before it is biologically necessary. | A Bayesian computational model of language learning in construction grammars can learn a language, in a bootstrap starting from no language knowledge.<br><br>Bayesian language learning is robust against some failures to understand correctly what the speaker is referring to in the common ground; it only needs some successes.<br><br>The Bayesian learning process can learn thousands of constructions.<br><br>The mathematical basis of the learning guarantee is described in the appendix. | Simple constructions defining core language are needed in any language, so that children can start learning early; these constructions survive the evolutionary selection of constructions. |
| Any language has a learning guarantee – that all people in a speaking community will learn some subset of the same set of constructions | If the people in a speaking community could not learn highly overlapping sets of constructions, there would be no basis for the communication guarantee, so conversations would not be viable and could not perform their selective function. | The learning guarantee can be proven mathematically from the properties of unification and the complementary operation (intersection of feature structures) which is used for learning.<br><br>The learning guarantee can be demonstrated in a working computational model. | |

| Conversations are cooperative and repairable | Both people in a conversation have an incentive to extend the conversation, to show off their intelligence – so they have an incentive to cooperate, and to repair conversations if needed. | The constructions of a language have a defined nested turn structure, to help each person understand the structure of the conversation, and the current place in it.<br><br>The communication guarantee enables cooperative conversations, without prohibitive amounts of repair. | Any pragmatic constructions in a language, which undermined the turn structure of its conversations, would not survive the evolution of constructions. |
|---|---|---|---|
| Languages include a wide range of pragmatic devices | Each pragmatic device is a test of intelligence and speed; so they proliferate, to make the competition harder. | Languages contain large numbers of learned pragmatic constructions, which are applied by fast unification. These handle all pragmatic devices, such as presupposition, implicature, speech acts, irony and metaphor.<br><br>Most pragmatic devices rely on profligate learning of constructions, rather than on any elegant logical analysis.<br><br>Pragmatics and semantics overlap seamlessly in constructions, and in the sequences of unification for language use.<br><br>The same pragmatic constructions are unified in different orders by speakers and listeners, underpinning the communications guarantee | Pragmatic constructions are subject to strong selection pressures, depending on their usefulness in the community, potential for ambiguity, economy, learnability, and so on. They evolve rapidly in a speaking community. |
| Every language is a mixture of regularity and irregularity | Our minds have evolved to understand and use a language, however irregular it may be; but some regularity makes language learning quicker and easier.<br><br>While elegant syntactic accounts of language are intellectually satisfying, concise syntax is not necessary; we can learn many thousands of constructions. | The computational model of construction grammars can handle any degree of irregularity; but regularity simplifies the parsing of complex utterances, and handling ambiguities; and makes language learning faster. | In the evolution of constructions, frequently used constructions may be highly irregular (as there are many learning examples for them); however, less frequently used constructions need the advantage of easy learnability (by regularity - good alignment with other constructions) in order to reproduce and survive |
| Conversations are multi-modal – involving voice, gesture, facial expression and so on | The precursors of language in the intention game were multi-modal, and conversations have remained multi-modal.<br><br>Multiple channels enable us to convey information fast, in order to be impressive conversationalists. | In construction grammars, the constructions can contain multi-modal content, with constraints on relative timing, in their 'phonetic' poles.<br><br>The language learning procedure propagates multi-modal constructions from one generation to the next. | |
| People start their next turn in a conversation after a very short delay – typically 200 milliseconds | There is intense selective competition to respond fast – both to show off impressive intelligence, and to get into a conversation before someone else does. | When understanding an utterance, unification is done fast and in parallel, building partial meanings (such as speakers' broad intentions) early, to start planning a response. | If a construction does not support early resolution of the conversational turn type and speaker's intention, then it will not be used as much, and will fail to reproduce in the speaking community. |

This table shows that to understand any language phenomenon, it may be necessary to combine evolutionary, computational and social/historic viewpoints. This theory of language is a framework for doing that. I am not aware of any major language phenomenon that cannot be accounted for in this way; but language is so vast, that much detailed comparison work remains to be done.

Meanwhile I suggest that, as the table shows, the unified theory of language has a positive Bayesian balance between the complexity of its hypotheses (on one side of the balance) and the number of facts about language that it accounts for (on the other side). The facts it accounts for outweigh the complexity of its hypotheses. A positive Bayesian balance is the criterion used in science to recognize a successful theory, as one which has a high posterior probably of being correct, in the light of the evidence [Sprenger & Hartmann 2019]. This implies that the unified theory of language is on the right track, and merits further investigation.

## 16. Language and Human Nature

If, as suggested in the previous section, the unified theory of language is on the right track, then it can tell us something new about human nature, giving us a fresh viewpoint on ourselves and our lives. It implies that language is the origin of the huge differences between human beings and other animals; that language is responsible for most of our human nature.

Exploring human nature through the lens of language is the subject of other papers [Worden 2024a, 2024b, 2025]. This section is a brief introduction to those ideas. The picture of human nature which emerges may initially seem to be a dark one (and you may feel some emotional resistance to it); but the full picture has great positive potential; so please read to the end of the section.

Language is commonly regarded as a kind of neutral medium for the expression of truths about the world; as a vehicle for rational thought, through logic, reasoning and argument. We trust language, as a fish trusts water. Our minds swim in language; how could it deceive us? As will be described below, it does. If we understand its evolutionary history, as a tool we use to compete for status in social groups, in order to pass on our genes, then we should not be so trusting of language.

Language is primarily a tool for assessing social status, and only secondarily a tool for expressing truths about the world. Many properties of today's language support this view – that language is not neutral, but imports and biases and fictions to our thinking, expressly for the purpose of computing social status.

A primary aspect of language (in its grammatical structure) is the structure of a declarative utterance, which is agent-act-patient or agent-act (i.e. subject-act or subject-act-object). Why is there always an obligatory agent/subject? Why is there some entity (often a person) who language makes responsible for the act? The reason is that by making people responsible for acts, you can assign blame or praise for the act or its consequences, and assign social status points to people. We do this at many moments in the day. When thinking, we often infer our own status, in the eyes of some imagined 'shadow audience' which represents how we think our social group will regard our status. Once you have noticed these core status-computing biases in our language and thoughts, other biases can be found:

| Aspect of language (in speech and thought) | Why the aspect is a fiction (why it does not describe reality) | Purpose it serves computing social status, in the competitive status game |
|---|---|---|
| Agency (grammatical structure of utterances is agent-act-patient or agent-act, i.e subject-verb) | Agency does not exist in nature. Things and animals do not choose to do things; events just happen.<br><br>Agency and free will are not a true innovation of the human mind. | By making people responsible for events, language assigns blame and praise (social status points) to people. |
| self | The human agent self is defined by our mental shadow audiences: 'how I think other people regard me and my status'. This second-hand view of ourselves and our own status is frequently incorrect.<br><br>Other people spend little time thinking about our status. Most of their time is spent thinking about their own status. | Our estimate of how other people regard us is self-esteem; this estimate of our own status drives most of our emotions and behaviour, making us play the status game. |
| I/me (and other pronouns) | 'I' and 'me' conflate two separate concepts:<br>1. the agent self (who acts and scores status points)<br>2. the experiencing self (who does not act, but consciously experiences what happens to it).<br><br>In reality, the two are different. This dualism arises from language (and is refuted in Eastern philosophy) | We necessarily confuse these two concepts, to make us play the status game seriously.<br><br>If we did not, we might just opt out of the game, and human society would not hold together. |
| Causation | Reality is not a time-ordered network of causal links; it is a pattern in the four dimensions of space and time, obeying mathematical constraints (determinate or random), as described by science.<br><br>Causality an optional way to describe nature, in certain approximations. Better descriptions of nature do not rely on causality. | By saying 'person P does A; A causes B', language expands the number of events that P can be blamed or praised for. This makes the status game more all-encompassing. |
| Chronological time (future and past tenses) | We do not consciously experience anything except the present moment – a time slice of thickness about 1/5 second. Animals experience no extended time; they only experience the present moment.<br><br>The patterns of the physical world are described locally in time and space, by the relations between adjacent times, rather than by long-range relations across time.<br><br>Conscious language creates a distant past and future in our minds. | Past and future increase the number of ways to assign blame and praise to people. They also create intervals for causation to act in. This expands the scope of the status game.<br><br>Linguistic tense supports life stories, and the status evaluation of life stories. We have constant autobiographical anxiety, forcing us to compete to make a 'better' life story, with higher status.<br><br>Our anxiety includes fear of the life story ending. |
| can | To say 'P can D' implies that P has a choice, to do D or not do D. In reality, D just happens, or not; it is part of the pattern of events in space and time. | Talking about what people 'can' do, even when they do not do it, creates more opportunities for blame and praise, assigning status. |

| Comparatives & alternatives | Alternatives (similar to 'can') are events that do not happen; so they do not exist. The world is as it is – a pattern in space and time. | Alternatives exist to be compared. Comparison creates relative displacements in a status space (like relative displacements in real space). This enables the discussion of relative status.  Comparisons across time are useful status talking points. |
|---|---|---|
| Negation (not) | Events that do not happen do not exist, in the spacetime pattern of reality.  (mathematically, there is an infinity of non-events which do not happen; they are not defined) | Negation enables a kind of comparison – comparing event E with not E, in terms of the status of some person. |
| Nested utterances for mental states (intentional stance) | Mental states like 'P intends to A' imply that people have choices, to act or not act. People do not have choices or free will; events just happen, involving people. | Attributing mental states, such as intentions, increases the number of ways we can assess a person's status – assigning blame and praise. |
| Language embodies a folk theory of the world (folk psychology, folk physics, folk medicine, etc.) | Each folk theory of the world is made up arbitrarily by people. It has no limits, and it is scarcely subjected to experimental checks; in this, it is not like a testable scientific theory.  Folk theory includes myths, for which there is no evidence. Different folk theories conflict with one another. | Folk theory is used by people to assign status points to other people, and to themselves. It defines the status values of events (good and evil). |

In this way, language spins a complex web of related concepts, which we use to calculate the social status of people, particularly ourselves. That is a simple descriptive fact about language today, which you can check from your own experience. The central column of the table says that many of these concepts are not parts of nature. You may find this a radical or unpalatable view of language, but please consider it carefully before rejecting any of it; it is supported by much Eastern philosophy and by modern science.

What evolution led us to suspect – that language is a biased tool for computing status, not a neutral tool to describe the world – is confirmed in the table by examining language as it is today. Evolutionary arguments are not necessary for the conclusion; they simply support the point.

The nature of language has consequences for the practice of philosophy. Many Western philosophers have put their trust in language, adopting it as their main intellectual tool. They have seen the tools of language – logic, reasoning and argument – as a kind of *Camino Real* to the truth, a concise shortcut to true knowledge. From the evolutionary view of the nature and purpose of language, it cannot be so trusted; there is no yellow brick road to the truth, just the hard slog of comparing ideas with evidence, as in the halting progress of science. Eastern philosophy has not trusted language, going back as far as Lao Tzu – 'the way that can be spoken of is not the true way'. Eastern philosophy starts from an empirical study of the mind. In a saying attributed to the Buddha: 'do not believe what I say; try it for yourself'. Much Western philosophy has value; but it may best be regarded as a set of signposts for empirical investigation, to be used with a proper mistrust of language, and not as any last word.

I next examine the micro-structure in time of language and thought.

Verbal trains of thought – like those which occupy most of our waking moments – are a use of language. There is a competitive reason why we are constantly thinking: we think in order to mentally rehearse for conversations, to make them as impressive as we can. Verbal thinking is a constant activity, and it is often concerned with issues of our status in our social groups – with our self-esteem, through the shadow audience.

When we have a verbal thought, we become consciously aware of it in three ways:

a) We hear the sounds of the words, in a voice located somewhere in our heads
b) We consciously form mental images of the meanings of the words and language fragments
c) We use fast unification, both to make a sequence of words, and to work out the implications of the thought for our own social status, in the eyes of a shadow audience ('If I said this, what would people think of me?'); from the resulting impact on our self-esteem, we feel social emotions, as conscious sensations in our bodies.

What is the fine structure of these processes in time, for a typical thought, that may take one or two seconds to think?

The process for producing language (both overt and internal) uses fast, pre-conscious unification of constructions. This happens in the fractions of a second before the words are sequenced and heard as sounds in the head, and before their meaning is consciously imagined. So the time ordering of the three conscious experiences of language is approximately (c), (a), (b).

In the pre-conscious unifications that we do when thinking, the primary unification process is to construct a fragmentary utterance, but a secondary process happens at the same time. Some of the constructions are pragmatic Theory of Mind constructions, used to infer how the audience will receive what you say – to check that they will not misunderstand you, and that they will be impressed by your social status, as implied by the utterance. These constructions lead you to pre-consciously infer how your audience will judge your social status, from what you plan to say ('what will they think or me?'). These inferences are an important part of planning speech; you want to get it right, and appear impressive.

In a conversation, these Theory of Mind constructions are applied using the real audience for what you are about to say. In mental rehearsal for conversation, the same constructions are applied with some imagined audience – the sort of audience you think you might say that thing to. This hypothetical audience in your head is called a shadow audience, and it is present for every thought you think. If you pause to reflect, you can often feel the presence of a shadow audience; you may even consciously imagine them. We constantly infer how a shadow audience would regard our own status, in the light of what we might say. The result is a nudge to our self-esteem, which we consciously feel in our bodies as an emotion.

The whole process of formulating the thought, and inferring how a shadow audience will regard our social status, happens by fast pre-conscious unification, in the fractions of a second before we are consciously aware of the meaning. Conscious imagination of the word sounds and of their meaning happens only after the pre-conscious unifications and nudges. That is why experience (c), the emotional nudge, comes before (a) and (b).

These pre-conscious emotional promptings can form a complex cascade, because as soon as we feel some negative emotion arising from lowered self-esteem, that leads to another pre-conscious reaction, which can be paraphrased: 'Some shadow audience can see me feeling this negative emotion; in their eyes, that is a low-status thing for me to do'. This leads to another small nudge of shame – and so on. These fast automatic cascades of emotion have been learned with language from an early age; a small child knows that people can see her crying. To try to head off the cascade, we try out some other conscious thought to boost our self-esteem; but by then it may be too late. These arbitrary cascades of negative emotion, large or small, are a fundamental part of our lives. They are always with us.

We should not be entirely negative about language-based thought. Self-esteem is what enables us to cooperate with each other as functioning members of society [Tomasello 2014], doing great things that no one person can do on their own. Language has made us able to cooperate, as no other animal can. As well as that, verbal thinking enables us to remember any thought as consciously heard words, and later recall it to build an extended train of thought (in a way that no other animal can) to solve difficult problems. In these ways, the whole of civilization owes its existence to language-based thought.

Language has other consequences for our understanding of human nature. For instance:

- Language controls the way we think about ourselves – our social agent self, rather than our physical body. Our primary way of think about the social self is not 'my self I think about me' but 'myself as I think other people think about me' – as defined by our learned Theory of Mind constructions. This is an impoverished, second-hand and distorted view of our selves as seen by our shadow audiences – like seeing ourselves in a cracked mirror
- Part of our self-esteem is a language-induced illusion that we are rational agents, making rational choices – whereas most of our choices are driven by fast, pre-conscious emotional reactions; then we retrospectively rationalize our motives [Chater 2022].
- We use language largely as tool to persuade people of our own point of view (and thus to increase our own status), rather than as a tool for analysing the true state of affairs [Mercier & Sperber 2017]
- Our self-esteem impels us to hold opinions just because we think they will agree with the opinions of other people in the group; not because they fit the evidence. Agreeing with other people is a way to gain status in the group. This is group-think [Worden 2024a, 2024b].
- As part of group-think, we believe that our group is somehow superior to other groups. Outsiders can be labelled in many different ways; their opinions, lives and happiness can then be ignored. This is tribalism. [Worden 2024a, 2024b]
- Part of our self-esteem is continually to maintain and burnish our life-stories, as we would like them to be – or rather, as we would like other people to think they are. These fanciful life stories are our constant anxious preoccupations and motivators.

The resulting viewpoint is an uncomfortable picture of human nature – a picture that you may want to rebut. While we all know that we have shadow audiences, group-think, tribal instincts, fanciful life stories and hidden irrational motives, we would like to think they do not matter much; that our rationality overrides them.

It is self-esteem which drives us to reject this darker picture of ourselves. Yet if we can step back from our self-esteem for a moment, this picture of human nature is not unrealistic. Understanding the dark side of human nature could be the first step to doing something about it. As in the history of medicine, we need to understand the disease before we can cure it. Understanding the role of language in our lives, and how it can harm us, can we start to cure some of its effects?

Looking at the sub-second micro-structure of thinking, the prospects do not seem good. Conscious rational thought is too slow to forestall the irrational emotional promptings which drive our actions. Emotion moves faster than thought, and forestalls thought. Yet there is a possible way forward.

In their beliefs about gods and afterlife, the world's religions disagree with one another – showing that most of them must be wrong. However, in their descriptions of a few core personal spiritual experiences, there is remarkable convergence. There is a shared mystical core to all religions. This core is described most clearly in the Eastern religions such as Buddhism, and is now commonly called mindfulness, linked with the practice of meditation. It is called 'mystical' for a simple reason. Because mindfulness is a private personal experience, it is not a part of the shared common ground which is the basis of language. Mindful experiences are difficult, if not impossible, to describe in language. They are ineffable and therefore mystical.

There is a wide range of techniques and practices to bring about a mindful state; I shall not discuss them much here. I focus instead on descriptions of mindful states, such as those in Metzinger 2024]. Mindful states are rarely described in positive terms of what they are, more commonly in negative terms, saying what they are not. Even when words like 'peaceful' are used, 'peace' can be seen as the absence of mental strife.

I note here a common factor, which seems to unite many of the things that are reported to be absent in mindful states: the absent things are parts of the pre-conscious micro-structure of language in the mind. Mindful states are reported in terms of:

a) Peace, or the absence of stress and anxiety (which can be taken to be fast language-driven cascades of negative emotions)
b) The absence of self (which can be taken to be the human social self, as defined by shadow audiences); no subject/object distinction, just non-dual experience.
c) The absence of time (absence of language-based extended time, and life stories)
d) Lightness, or the absence of some 'heavy' localized bodily feelings, as induced by language-based cascades of emotion
e) Spaciousness, or the absence of obstacles in a conscious space; perceived obstacles may be related to language-induced emotional feelings.
f) The absence of boundaries between self and the world.
g) The absence of persistent verbal thoughts – or when thoughts arise, they pass by with no reaction.
h) The absence of classifications of things into types, purposes, meanings or labels. This is sometimes called 'suchness'.
i) The absence of goals, and of attraction to ('clinging to') or of repulsion from things and states
j) Compassion, or the absence of judgement and mental rejection of other people (which can be taken as the absence of attempts to assess their social status, or create out-groups)

These absences do not include all the feelings reported in mindful states, but they describe a good part of them. What remains (what is not absent) is universally described in positive emotional terms. It may be that to compensate us for the emotional burdens brought by language (to re-balance the positive and negatives in our lives), recent evolution has given us emotionally positive phenomenal states, which we can experience when language and its effects are stilled.

Much of the central column in Table 2 above is reduced or absent in mindful states. This leads to the hypothesis:

> **Mindful states are states in which the activity of language in the mind is reduced or stopped.**

This hypothesis gives an economical account of the many reports of mindful states in [Metzinger 2024], as well as the personal experience of mindful states. It is a simple hypothesis, and accounts for much of the evidence. This means that it has a large Bayesian probability of being correct [Sprenger & Hartmann 2019].

We can now frame a hypothesis about how people experience mindful states, at any stage in their lives. Two important parts of our language heritage are the agent self, and the shadow audience. Neither of these are real, but we learn them both as parts of language from an early age. We may learn hundreds or thousands of language constructions in which some other person (the shadow audience) assesses the status of our agent self - and does so negatively. These constructions stay with us for all our lives; like common words, they are not un-learned. The agent self and the shadow audience, while both fictitious, continually re-create one another in our minds, in the vicious cycle of our fast cascades of emotion. This is the Buddhist cycle of birth and re-birth; the agent self is constantly re-born in our minds.

How does mindfulness change this? Mindfulness training – of whatever form – may involve language-like learning of new constructions, which focus more on physical awareness of feelings in the body, less on the agent self and what it might do next. Constructions compete, in the pre-conscious mind. The new constructions displace or inhibit the effects of the old agent self/shadow audience constructions. Mindfulness happens when we are not trying to be mindful

– when the agent self is no longer re-born, and there is no subject-object duality.

If that is a correct scientific understanding of mindful states and how they occur, it could in time lead to wider general understanding of what mindfulness is, and to wider adoption of mindful practices. This would give widespread benefits to humanity, at both a personal and social level. It might be the most important single consequence of this theory of language.

## 17. Conclusions

This paper has described a theory of language which aims to account for all the main empirical evidence about language, in a single unified framework. This is an ambitious aim; are there grounds to believe that the theory succeeds? Does the theory fit the data?

Comparisons with the evidence about language are summarized in section 15, in a table with four columns. This tabular form is used because the account of any language phenomenon depends on the three dimensions of the theory: evolutionary, computational, and historic. Each dimension has aspects that may be counter-intuitive:

1. **Evolutionary**: Researchers in any subject like to believe that knowledge of their subject will benefit mankind. This is a noble motive for many research careers. It is commonly assumed that language is a wholly benevolent adaptation of the human species – a neutral medium for rational thought, which sets us above all other animals. The theory that language evolved by sexual selection challenges this view. It implies that language is not primarily a tool for rational thought, but is primarily a tool we use to compete for social status and for mating chances – to pass on our genes. This less idealistic view of language runs through the theory in its account of many language phenomena, and it may be uncomfortable to some; but it fits the evidence of language as it is today.
2. **Computational**: The origins of the study of language lie in philosophy – whose main intellectual tool has been language itself. Accounts of language phenomena have striven for logical economy and elegance; looking to explain language phenomena by logic, reasoning and argument. In that spirit, some readers may prefer a Chomskyan theory of language, to the cognitive linguistic approach of this paper. Even if they lean towards cognitive linguistics, readers may prefer their computational accounts to have a crisp logical character – rather than Bayesian probabilistic accounts. In this theory, both animal and human brains are prodigious probability calculators, doing fast Bayesian pattern matching in ways that may not fit any simple logical account. This kind of explanation may be culturally unfamiliar to some readers; but again, the true test is the fit to the evidence.
3. **Historical**: Students of language like to think that it is telling us important truths about the human mind - truths of timeless importance. Language universals may be thought to reveal deep constraints on how the mind processes language. In this theory, much of the structure of language – including language universals – tells us less about the human mind, and more about recent historical processes of language change. This is a double evolution – evolution of the human mind, and evolution of the constructions of language. The second process is much faster than the first – and so it determines much of the outcome. Historic language change, rather than evolution of the human mind, is responsible for many prominent features of language. This may disappoint some readers; but again, the test is fit to the data of language.

So there are possible reasons to feel uncomfortable about the theory. I ask you to set aside any such discomfort and ask: does the theory fit the data? That is the ultimate criterion by which a scientific theory is judged.

## Appendix: Mathematical Basis of the Two Guarantees

An important part of this theory is a pair of mathematical guarantees – called the communication guarantee, and the learning guarantee – which together guarantee that language can do its job, of supporting successful communication.

The purpose of this appendix is not to give a full mathematical derivation of those guarantees, but to describe the underlying mathematics, and to outline the derivations.

The **communication guarantee** shows that, provided a speaker and a listener share the same set of language constructions, then the listener can, from words that he or she hears, reliably reproduce the meaning that the speaker used to produce the words.

The **learning guarantee** shows that, if a learner hears a small number of spoken learning examples, and can sometimes partially understands them, where some speaker describes something in the common ground, with all the examples using only one construction which the listener has not already learnt (and the other constructions are known to the listener), then the listener can use those examples to learn the new construction, in the same form as was used by the speakers.

The learning guarantee shows that a set of language constructions will propagate accurately across a speaking community, so that all people share the same set of constructions. This underpins the communication guarantee.

They are both statistical guarantees, showing that these things happen in the great majority of cases, in spite of occasional mis-hearing, misunderstanding, and so on. They rely on the Bayesian nature of language cognition. I first describe the mathematics of constructions and their operations.

Each construction in a language is represented by a feature structure, which is a directed acyclic graph (DAG) with property values, such as 'gender:male', defined on slots on its nodes. Variable values of properties are denoted by single upper-case letters, such as 'number:X'. If a variable X appears more than once in a DAG, it means that the value of X is not known, but X must take the same value everywhere it appears. X can denote anything – even a sub-graph. There can be time ordering constraints between the nodes of a graph.

The feature structure for a construction has two main branches from its root node, which are called its poles. It has a semantic pole, which consists only of meaning slots in a single sub-graph, and no sound nodes; and a 'phonetic' pole which may contain one or more sub-graphs. At least one of these sub-graphs contains a word sound slot, such as 'sound:'fred'', other sub-graphs in the phonetic pole hold only meaning slots. Typically, the phonetic poles of noun constructions contain only sound slots, whereas the constructions for other parts of speech contain other meaning parts in their phonetic pole.

Every meaning part in the phonetic pole of a construction is a typed variable – typed, in that it obeys a set of constraints. Each such variable occurs as a part of the semantic pole of the same construction, typically nested inside a large feature structure. These shared variables are the way in which complex meanings are broken into smaller parts (to express them in words) or combined (to understand them).

Feature structures (including constructions) have a model-theoretic semantics. Every feature structure F represents a set of states of affairs in the world – states of affairs that are constrained by its slot values. This set of states is called its **scope** σ(F).

Scope sets are used to define the fundamental relation between feature structures – the relation of **subsumption**. For two feature structures A and B, A subsumes B (written as A > B) if and only if every state of affairs in the scope σ(B) is also in the scope σ(A); that is, if σ(B) is a subset of σ(A) (written here as σ(A) > σ(B)):

$$A > B \text{ iff } \sigma(A) > \sigma(B)$$

This is the core relation between feature structures and set theory.

Feature structure A can only subsume B, if B includes all the graph structure of A, and all the same slot values at the same places. Otherwise, some states in σ(B) will not be in σ(A). B has to be a more complex version of A. Equivalently, B is A with some added extra detail.

The information content of a feature structure A – written as I(A) – is approximately the sum of the information content in all its slot values. If a slot has two possible values of equal probability, then fixing the slot to one of those values gives it an information content of 1 bit.

If A subsumes B, then B has more defined slot values than A. It therefore follows that

$$\text{if } A > B, \text{ then } I(B) > I(A)$$

We can also define the probability of a feature structure P(A), as the probability that any state of affairs is in the scope σ(A). This probability decreases as the information content of A increases (as more slots are added to A, a state becomes less likely to be consistent with A; is scope gets smaller). There is an approximate relation:

$$P(A) = 2^{-I(A)}$$

We can now define the two fundamental operations of feature structures, which are used for language processing and learning.

The **unification** of two feature structures is written as C = (A U B). C is defined by

$$A > C \text{ and } B > C$$

Or equivalently

$$\sigma(A) > \sigma(C) \text{ and } \sigma(B) > \sigma(C)$$

If the scopes σ(A) and σ(B) have no overlap (for instance, if some slot values in A are inconsistent with the values of the same slots in B), then the scope σ(C) is empty, and the unification C does not exist.

From the definition it follows that if C exists:

$$I(C) > I(A) \text{ and } I(C) > I(B)$$

In other words, C must have all the structure and slot values of A, and all the structure and slot values of B. C must be a larger structure than both A and B, containing them both as sub-structures. C is defined to be the simplest such feature structure – the one with the least information content (and therefore the largest probability P(C)) – which has this property.

This is the fundamental link between feature structure unification and Bayesian inference. If C = (A U B), The set of states σ(C) has the largest probability P(C) which is consistent with both A and B.

To compute the unification C = (A U B) from A and B, it is necessary to match the nodes of A and B together, respecting their structures and matching their slot values, finding the maximum possible match in information terms. This requires some trial and error, and it results in the smallest possible structure C which has both A and B as substructures.

As a result, unification is commutative

(A U B) = (B U A)

But not associative:

(A U (B U C)) ≠ ((A U B) U C)

In some cases, the first order of unification gives no result (e.g. B and C have no overlap), but the second order does (A and C have some overlap).

The other operation on feature structures, complementary to unification, is the operation of **intersection**, written as D = A ∩ B. The intersection D satisfies

D > A and D > B

and D is the most complex feature structure that satisfies these conditions. The scopes σ(A) and σ(B) need not overlap for D to exist.

To compute the intersection D = (A ∩ B) from A and B, you again try to match the nodes of A and B, respecting their slot values, looking for as much matching as possible, in a trial and error process to maximise the information content of D. If some slot has different values on a matching node of A and B, or is missing on one of A or B, it is missing from the result D. Parts of A and B which do not match are not included in D. So intersection is a feature structure shrinking operation; D is a smaller structure than both A and B, and has less information content:

I(D) < I(A) and I(D) < I(B)

Feature structure intersection is both commutative

(A ∩ B) = (B ∩ A)

and associative:

(A ∩ (B ∩ C)) = ((A ∩ B) ∩ C)

From the close relationship between feature structure operations and set theory, there are many algebraic relations between feature structure operations, which resemble the relations of set theory. For instance:

If (A U B) > X, then A > X and B > X

if A > X and B > X, then (A ∩ B) > X

I next outline how unification and intersection are used in language processing and learning, and how they lead to the two guarantees.

The recursive procedure for producing an utterance is given by the steps S1..S7 described in section 10, and the recursive procedure to understand an utterance is given by the steps L1..L5. I shall not repeat those steps here (this is not a detailed derivation of the communication guarantee); I shall only outline how the communication guarantee can be derived.

For a complex utterance with meaning feature structure M, both speaker and listener unify the same set of constructions, say A, B, C, D; and they also unify the same facts F, G, H.. from the common ground. But they unify them in different orders, and in different ways:

- The speaker starts from M and unifies in a nouns-last order, by matching the semantic pole of each construction, adding the phonetic pole to the growing feature structure. This carries on until unification has matched all meaning parts in the phonetic poles, adding word sound parts. The word sounds are then spoken.
- The listener starts from the word sounds and unifies in a nouns-first order, by matching the phonetic pole (including word sounds), adding the semantic pole to the growing feature structure. This carries on until unification has matched all the word sound parts of the heard utterance.
- Both speaker and listener 'weave in' unifications to the facts F, G, H from the common ground, in their different unification orders.

If the speaker does binary unifications in a certain associative order [such as ((A U F) U B)…], to build a large feature structure, which we call Σ. contains all the feature structures A, B, C, F, G, H.. as substructures (each of them subsumes Σ). Σ contains both the original meaning structure M, and the word sounds of the constructions. The speaker then speaks the words, which the listener hears. The listener can unify the same feature structures in exactly the reverse order [an order such as ((B U F) U A)…] , creating another large feature structure Σ'.

(if the phonetic poles of constructions are envisaged on the left, with the semantic poles on the right, in the speaker's unification process, information such as slot values flows from right to left; in the listener's unification process, the same information flows from left to right)

Σ and Σ' contain the same set of sub-structures, with the same slot values, so it can be shown that they are equal.

Σ' = Σ

Therefore Σ' contains the same original meaning structure M, that the speaker started from. The listener can reconstruct the speaker's original meaning. That is why the communication guarantee works. Working examples of this can be examined in detail in an online demonstration at [Worden 2022] (the examples do not have common ground feature structures, but the principles are unchanged).

For this I shall use a tabular notation, to show the sequence of unifications done by the speaker (in his version of the common ground), and the unifications done by the listener

(in her version of the common ground). The example is the sentence 'he sleeps', referring to a male person (harry) who is visible to both the speaker and the listener – who is in the common ground. The speaker and the listener use the same two language constructions, which are:

(C1)    ['he'|MALE S]

This construction embodies the language fact that the sound 'he' (its phonetic pole) denotes a male, singular person (its semantic pole).

(C2)    [ T |'sleeps' | [T] SLEEP PRESENT ]

This construction embodies the language fact that some thing T (which is constrained to be male and singular; the constraint is not shown explicitly), followed by the sound 'sleeps', denotes a state in which the thing T is in the act of sleeping at the present moment. Its phonetic pole has two parts (T and 'sleeps'), while its semantic pole has one scene denoting a thing T sleeping now. The variable T occurs in both poles of the construction, and acts as a conduit for meaning to move between the poles in unifications. Square brackets [] denote nesting of nodes in feature structures. Vertical bars ('|') denote separate top-level branches of a feature structure.

The speaker wishes to express the meaning [[HARRY] SLEEP ] . This meaning, we assume, is not yet in the listener's common ground. (for simplicity, the tense PRESENT has been left out of the example) The speaker wishes to add word sounds to the common ground, so that the listener can infer this meaning. The unifications done by the speaker are shown in the first table:

| Row | A | B | R = A U B |
|---|---|---|---|
| 1 | [ T |'sleeps' | **[T] SLEEP** ] | **[[HARRY] SLEEP** ] | [ HARRY |'sleeps' | [HARRY] SLEEP ] |
| 2 | [**HARRY** MALE S]    (CG) | R1 | [ HARRY MALE S |'sleeps' | [HARRY MALE S] SLEEP ] |
| 3 | ['h**e**'|**MALE S**] | R2 | [ 'he' | HARRY MALE S |'sleeps' | [HARRY MALE S] SLEEP ] |

In row 1, the speaker unifies his intended meaning (in column B, cell B1) with the language construction for 'sleeps' (in column A, cell A1). The parts of A and B which match in the unification are shown in **bold**. The construction has been unified from right to left, matching its semantic pole. In the result column R (cell R1), the shared variable T in the construction has been fixed to the value HARRY, creating a new standalone meaning HARRY (on the left-hand side of R1), which the speaker needs to express. R1 also contains the sound 'sleeps', which will later be spoken.

In row 2, the speaker does a prominence-first search of the common ground, for facts which will enable him to concisely express the meaning HARRY. This search is as described in section 11. He finds that the person HARRY is the most prominent single male in the common ground; so he unifies this fact (shown in cell A2) with the result R1 of the previous unification (now shown in cell B2), matching the meaning HARRY. The result in cell R2 is an enrichment of R1, recording that HARRY is male and singular.

In row 3, the speaker unifies the construction for the word 'he' (shown in cell A3) with the previous result R2, now shown in cell B3 – again, matching on the semantic pole of the construction. The result R3 now contains the sound 'he', as well as the sound 'sleeps'. No further meanings need to be expressed, so there are no further unifications by the speaker – who then says the words 'he sleeps', in the time order defined by R3.

The unifications done by the listener are shown in the next table.

| Row | A | B | R = A U B |
|---|---|---|---|
| 1 | ['**he**'|MALE S] | ['**he**'|'sleeps'] | ['he'| MALE S |'sleeps'] |
| 2 | [HARRY **MALE S**]    (CG) | R1 | ['he'| HARRY MALE S |'sleeps'] |
| 3 | [ T |'**sleeps'** | [T] SLEEP ] | R2 | ['he'| HARRY MALE S |'sleeps'| [HARRY MALE S] SLEEP]] |

The listener has heard the words 'he sleeps', so they are in her common ground (cell B1). The listener proceeds to unify constructions in a nouns-first order, matching their phonetic poles, until all word sounds have been matched.

In row 1, the listener matches the construction for 'he' by its phonetic pole, producing the result shown in cell R1.

In row 2, the listener searches her common ground for the most prominent male singular thing, and finds the fact shown in cell A2. This fact is unified with the result R1 of the previous step (now put in cell B2), giving the result in cell R2.

In row 3, the remaining sound 'sleeps' is matched with the phonetic pole of a construction. The only matching construction is that shown in cell A3, and it is matched with the result R2 of the previous step (now shown in cell B3). In this unification, the shared variable T in the construction has been fixed, to move the meaning HARRY MALE S from left to right, so that it ends up in the right-hand end of R3. The listener has understood the speaker's intended meaning – as in the communication guarantee.

Three points are illustrated by this example:

- The speaker and the listener unify the same set of feature structures (constructions, or other feature structures in the common ground) in different orders of unification; they each build up the same large feature structure, shown in cell R3 of both tables. This is the essence of the communication guarantee.
- Pragmatic operations (finding and unifying facts in the common ground) are interleaved with syntactic/semantic operations (unifying language constructions). There is no pragmatics/semantics divide.
- Although the example only uses two constructions, the same principles apply to more complex utterances with different pragmatic devices, with more complex syntax and semantics. The proof of this is a simple recursion.

I next describe the learning guarantee.

Consider a child who has learnt constructions A, B, C, D, but who has not yet learnt some construction X. Adults around her carry on conversations, sometimes using the words A, B, C, D and X. She can recognize which utterances contain the unknown word X, because she hears (but does not understand) its novel sound part. She also observes, and partially understands, facts in the common ground. For those utterances which only use the words A, B, C, D, which she knows, she can understand what is said, using the listener's unification process outlined above, and can match the meaning M to some fact in the common ground.

Now consider a set of utterances which contain the words A, B, C, D and all contain X. The child gathers a few of these as **learning examples**. For each such example, she can partially understand it, doing those parts of the listener's unification procedure which only require A, B, C, D, (and facts F, G, .. in the common ground), and which do not use the unknown construction X. She can use this partial meaning feature structure to make a highly constrained guess as to which meaning M in the common ground is being referred to. (Very few of the most prominent meanings M will be subsumed by the partial meanings she has understood, so they will not match it) On some occasions this guess will be correct. In those cases, she can do the part of the speaker's production process (how would I express the meaning M?) which do not depend on X – checking that the words which she has heard are used to express this meaning.

In summary, for each learning example $L_i$ (i = 1..N), the child can do some of the listener's left-to-right unification processes, then do some of the speaker's right-to-left unification processes from a guessed meaning $M_i$, to form an **enriched learning example**, a large feature structure denoted by $\Xi_i$. She forms the intersection of all the enriched learning examples:

$$\Omega = \Xi_1 \cap \Xi_2 \cap \Xi_3 \cap .. \Xi_N$$

(the intersections can be in any order, as intersection is associative)

We can relate the child's enriched learning examples $\Xi_i$ to the large feature structures $\Sigma_i$ used by the speakers to produce each utterance. We can show that

$$\Sigma_i > \Xi_i \quad \text{for all i}$$

($\Xi_i$ may contain more information than $\Sigma_i$, because the child may have noticed a part of the meaning $M_i$ which the speaker did not express, and put it in her $\Xi_i$)

Because each $\Sigma_i$ is the result of a set of unifications, including the unknown construction X, we know that

$$X > \Sigma_i$$

Because the subsumption relation is transitive:

$$X > \Xi_i \quad \text{for all i}$$

which in turn implies that

$$X > \Omega$$

The result $\Omega$ of the multiple intersection process contains the unknown construction X as a sub-structure – and it may contain extra structure as well.

We can illustrate the intersection process for a child learning the construction for 'sleeps', having learn the constructions for some nouns, such as 'man', 'he' and 'Anna', which appear in three learning examples. In each case, the child correctly infers the intended meaning, and constructs the enriched learning examples:

| Learning example | Enriched learning example $\Xi_i$ |
|---|---|
| A = 'man sleeps' | ['man'\| M S \|'sleeps'\| [ M S] SLEEP]] |
| B = 'he sleeps' | ['he'\| M S \|'sleeps'\| [ M S] SLEEP]] |
| C = 'Anna sleeps' | ['Anna'\| ANNA F S \|'sleeps'\|[ANNA F S] SLEEP]] |
| $\Omega$ = A $\cap$ B $\cap$ C | [$T_s$ \|'sleeps' \| [$T_s$] SLEEP ] |

Here, M is short for MALE and F is short for FEMALE. The tense PRESENT has again been omitted for brevity. The intersection operation, which is used to find Ω, has the following properties:

- It is associative, so the order of learning examples does not matter.
- Non-matching word sounds like 'he','man' and 'Anna' are simply removed from the result.
- Intersection discovers the repeated constrained variable $T_s$, which is constrained to be a singular thing. (in this example, $T_s$ is found by $T_s$ = (M S) ∩ (M S) ∩ (ANNA F S) = S
- Intersection can discover timing constraints between parts of the phonetic pole, when they hold in all the intersected examples; or can discover unordered constructions without timing constraints

As more learning examples are used, the intersection result Ω converges closer and closer to the correct construction for 'sleeps'. This is the learning guarantee.

As described above, repeating constrained variables like $T_s$ (in both the phonetic and the semantic pole of the construction) are the key to building up or breaking apart complex meanings. A complex construction can have several repeated variables, each denoting a thing or an event.

For any extra information not in X to remain in Ω, it must be present coincidentally in all of the enriched learning examples $\Xi_i$. As the number N of learning examples increases, the probability of any extra information being in all of them decreases exponentially and rapidly. Even after a small number of learning examples, Ω becomes a good approximation to the unknown construction X. This is the learning guarantee, and it is the reason why children learn words so fast, from only a few learning examples per word.

This is a high-level picture of the learning procedure. It has missed out some details, such as recognizing word segmentation, dealing with nested utterances, and making the procedure robust against misunderstood utterances – where the child picks the wrong meaning $M_i$ in the common ground. Bad learning examples can be filtered out of the intersection, because if they are intersected with the other examples, they would reduce the meaning information content of the learned construction X to near zero.

This learning procedure can learn a language in a bootstrap manner from a starting point of no vocabulary, including complex constructions with several shared variables. This can be seen working online in [Worden 2022a] (this demonstration does not include facts in the common ground; but the principles are unchanged).

Throughout this appendix I have referred to word sounds in the phonetic poles of constructions. The 'phonetic' poles can also contain gestures, facial expressions, vocal tones, and other facets of multi-modal communication, together with their timing constraints; the same mathematical guarantees go through, and they apply to pragmatic constructions including Theory of Mind constructions.